\documentclass[10pt,english,reprint,aip,cha]{revtex4-1}
\pdfoutput=1
\usepackage{mathptmx}
\usepackage{helvet}
\usepackage[T1]{fontenc}
\usepackage[latin9]{inputenc}
\usepackage{xcolor}
\usepackage{pdfcolmk}
\usepackage{babel}
\usepackage{refstyle}
\usepackage{amsmath}
\usepackage{graphicx}
\PassOptionsToPackage{normalem}{ulem}
\usepackage{ulem}
\usepackage[unicode=true,pdfusetitle,
 bookmarks=true,bookmarksnumbered=false,bookmarksopen=false,
 breaklinks=false,pdfborder={0 0 1},backref=false,colorlinks=false]
 {hyperref}

\makeatletter

\AtBeginDocument{\providecommand\figref[1]{\ref{fig:#1}}}
\AtBeginDocument{\providecommand\eqref[1]{\ref{eq:#1}}}
\RS@ifundefined{subref}
  {\def\RSsubtxt{section~}\newref{sub}{name = \RSsubtxt}}
  {}
\RS@ifundefined{thmref}
  {\def\RSthmtxt{theorem~}\newref{thm}{name = \RSthmtxt}}
  {}
\RS@ifundefined{lemref}
  {\def\RSlemtxt{lemma~}\newref{lem}{name = \RSlemtxt}}
  {}

\providecolor{lyxadded}{rgb}{0,0,1}
\providecolor{lyxdeleted}{rgb}{1,0,0}

\usepackage{enumitem}		

\usepackage{ifthen}
\renewenvironment{figure}[1][]{%
 \ifthenelse{\equal{#1}{}}{%
   \@float{figure}
 }{%
   \@float{figure}[#1]%
 }%
 \centering
}{%
 \end@float
}

\usepackage{caption}
\usepackage[compact]{titlesec}

\AtBeginDocument{\setlength\abovedisplayskip{1pt}}
\AtBeginDocument{\setlength\abovedisplayshortskip{1pt}}
\AtBeginDocument{\setlength\belowdisplayskip{1pt}}
\AtBeginDocument{\setlength\belowdisplayshortskip{1pt}}
\AtBeginDocument{\setlength\textfloatsep{6pt}}
\AtBeginDocument{\setlength\floatsep{6pt}}

\usepackage{babel}

\makeatother

\begin{document}

\title{Synchronization of pairwise-coupled, identical, relaxation oscillators
based on metal-insulator phase transition devices: A Model Study}

\author{Abhinav Parihar}

\email{aparihar6@gatech.edu}

\affiliation{School of Electrical and Computer Engineering, Georgia Institute
of Technology, Atlanta, Georgia 30332, USA}

\author{Nikhil Shukla}

\email{nss152@psu.edu}

\affiliation{Department of Electrical Engineering, Pennsylvania State University,
University Park, Pennsylvania 16802, USA}

\author{Suman Datta}

\email{sdatta@engr.psu.edu}

\affiliation{Department of Electrical Engineering, Pennsylvania State University,
University Park, Pennsylvania 16802, USA}

\author{Arijit Raychowdhury}

\email{arijit.raychowdhury@ece.gatech.edu}

\affiliation{School of Electrical and Computer Engineering, Georgia Institute
of Technology, Atlanta, Georgia 30332, USA}
\begin{abstract}
Computing with networks of synchronous oscillators has attracted wide-spread
attention as novel materials and device topologies have enabled realization
of compact, scalable and low-power coupled oscillatory systems. Of
particular interest are compact and low-power relaxation oscillators
that have been recently demonstrated using MIT (metal-insulator-transition)
devices using properties of correlated oxides. Further the computational
capability of pairwise coupled relaxation oscillators has also been
shown to outperform traditional Boolean digital logic circuits. This
paper presents an analysis of the dynamics and synchronization of
a system of two such identical coupled relaxation oscillators implemented
with MIT devices. We focus on two implementations of the oscillator:
(a) a D-D configuration where complementary MIT devices (D) are connected
in series to provide oscillations and (b) a D-R configuration where
it is composed of a resistor (R) in series with a voltage-triggered
state changing MIT device (D). The MIT device acts like a hysteresis
resistor with different resistances in the two different states. The
synchronization dynamics of such a system has been analyzed with purely
charge based coupling using a resistive ($R_{C}$) and a capacitive
($\ensuremath{C_{C}}$) element in parallel. It is shown that in a
D-D configuration symmetric, identical and capacitively coupled relaxation
oscillator system synchronizes to an anti-phase locking state, whereas
when coupled resistively the system locks in phase. Further, we demonstrate
that for certain range of values of $R_{C}$ and $C_{C}$, a bistable
system is possible which can have potential applications in associative
computing. In D-R configuration, we demonstrate the existence of rich
dynamics including non-monotonic flows and complex phase relationship
governed by the ratios of the coupling impedance. Finally, the developed
theoretical formulations have been shown to explain experimentally
measured waveforms of such pairwise coupled relaxation oscillators.
\end{abstract}

\maketitle
\begin{quotation}
\textbf{With insurmountable challenges facing silicon scaling, research
has started in earnest to identify potential computational architectures
and device technologies for a post-silicon era. One such paradigm
where coupled oscillatory systems perform computational tasks such
as pattern recognition and template matching has garnered recent interest;
and it necessitates the fabrication of compact and scalable oscillators
that can be electrically coupled. Recent advances in the development
of correlated oxides have led to successful demonstration of coupled
relaxation oscillators. Here we investigate the coupling dynamics
of pairwise coupled identical relaxation oscillators based on such
correlated materials that exhibit electrically controlled metal-insulator-metal
phase transitions. Using analytical and numerical techniques we show
how the coupling network can lead to in-phase and out-of-phase locking
as well as create a bistable system. Such systems have already been
shown to offer computational capabilities beyond the traditional Boolean
fabric.}
\end{quotation}

\section{Introduction}

Synchronization of systems of oscillators has attracted widespread
attention among physicists, mathematicians and neurobiologists alike.
Even simple descriptions of oscillators and their coupling mechanisms
give rise to rich dynamics. Synchronization dynamics of coupled oscillators
not only have a wide variety of applications in engineering \textbf{\cite{Datta:2014aa,Shukla:2014aa}}
but they also explain many natural, chemical and biological synchronization
phenomena like the synchronized flashing of fireflies, pacemaker cells
in the human heart, chemical oscillations, neural oscillations, and
laser arrays, to name a few \cite{Dorfler:2012aa}. Coupled sinusoidal
oscillators have been extensively studied \cite{Winfree:1967aa},
\cite{Kuramoto:1975aa,Kuramoto:2003aa} and their application in the
computational paradigm has been well demonstrated\cite{Nikonov:2013aa,Izhikevich:2000aa}.
A generalized description of oscillators in these models is usually
a canonical phase model \cite{Dorfler:2012aa,Mallada:2013aa}, and
the coupling mechanisms is generally assumed weak and composed of
simple periodic functions. Several studies on more general periodic
coupling functions have been studied \cite{Acebron:2005aa}. Along
with sinusoidal oscillators, non-linear Van-der-Pol oscillators and
several of its variants have also been studied and the applicability
of such models in neurobiological and chemical oscillators have been
demonstrated \cite{Rand:1980aa,Storti:1982aa,Kouda:1982aa,Chakraborty:1988aa}.
Such analytic models of coupled oscillatory systems almost always
require a canonical phase description of the oscillators and a periodic
phase dependent additive coupling that can be classified as weak.
Although such a description of a system of oscillators is elegant
and provide key insights, relaxation oscillators that have recently
been demonstrated using phase transition MIT devices, cannot be modeled
using such a simple phase description. Prior work by the authors have
experimentally demonstrated locking and synchronization in a pairwise
coupled system of relaxation oscillators \cite{Shukla:2014aa} and
its possible application in computation has also been discussed\cite{Datta:2014aa}.
The coupling behavior of relaxation oscillators illustrate complex
dynamical properties \cite{Saito:1988aa} and in this paper we study
the synchronization behavior of a pair of identical and electrically
coupled relaxation oscillators \cite{Shukla:2014aa}. Individual oscillators
are composed of either two MIT devices in series (D-D configuration)
or a MIT device in series with a linear resistor (D-R configuration)
\cite{Shukla:2014aa} and electrical coupling is enabled through a
parallel connected R-C network. We show, through analytical and numerical
techniques how the final steady state relative phase of such coupled
oscillators depend on the coupling function. For certain range of
values of the coupling function, we note the possibility of a bistable
system, where both in-phase and out-of-phase locking are stable, thereby
giving rise to the possibility of using such oscillatory networks
in computation \cite{Izhikevich:2000aa,Datta:2014aa}.

\section{Electrical Circuit Model and Representation}

\begin{figure}
\includegraphics[scale=0.7]{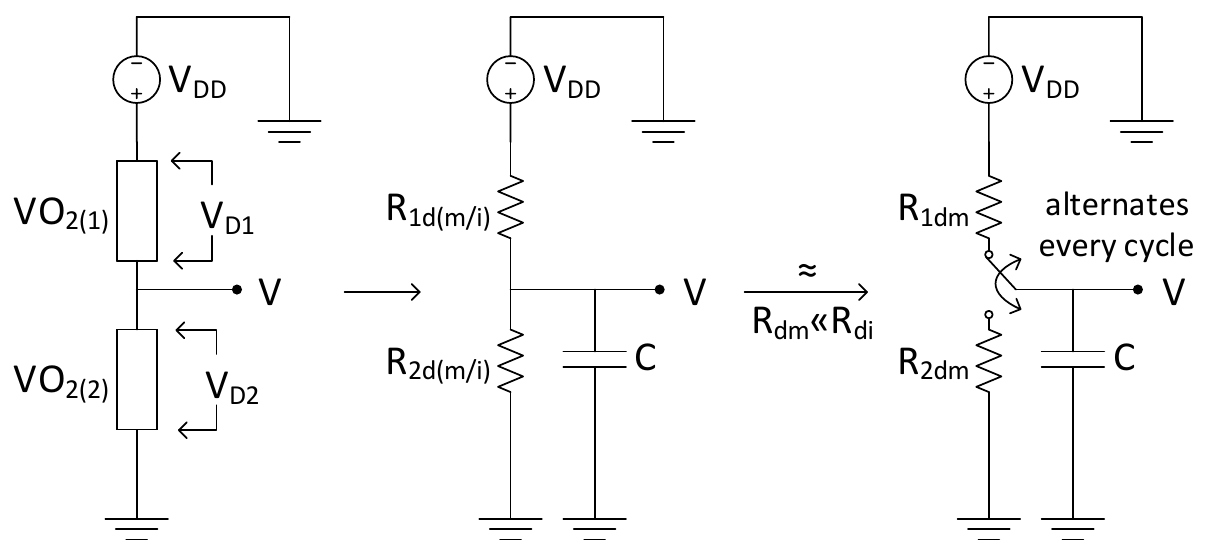}

\protect\caption{Relaxation oscillator circuit realized using two MIT state-changing
devices in series (D-D configuration), and its circuit equivalent
with $R_{dm}$ and $R_{di}$ as the internal resistance of the MIT
devices in metallic and insulating states respectively. When $R_{di}\gg R_{dm}$
the device behaves as a parallel combination of a capacitor and a
resistor with a switch.\label{fig:DD-circuit}}
\end{figure}

\begin{figure}
\includegraphics[scale=0.7]{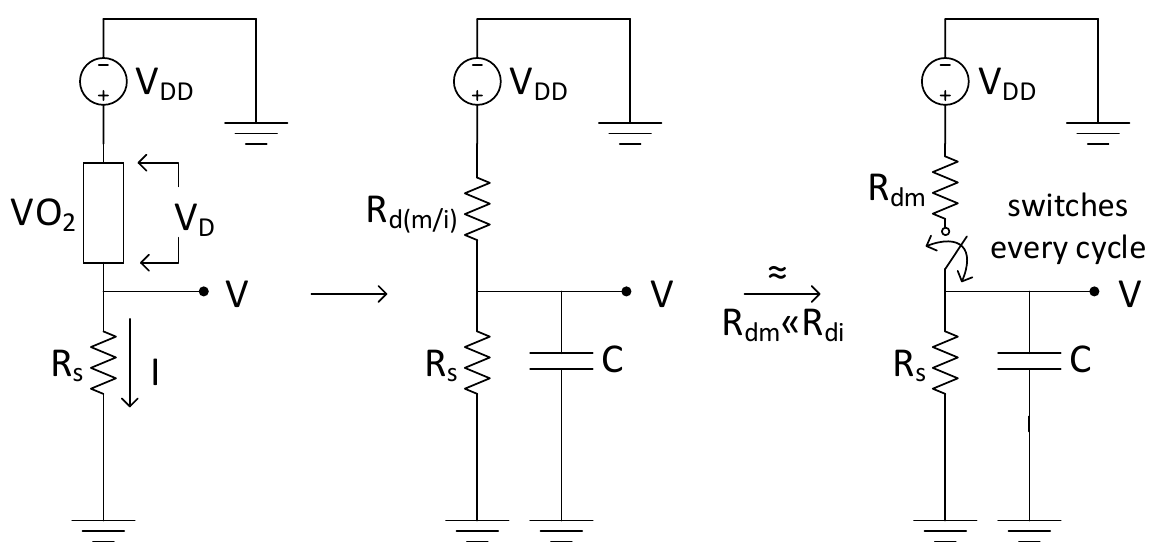}

\protect\caption{Relaxation oscillator circuit realized with a MIT device in series
with a resistor (D-R configuration), and its circuit equivalent with
$R_{dm}$ and $R_{di}$ as the internal resistance of the MIT device
in metallic and insulating states respectively. When $R_{di}\gg R_{dm}$
the device behaves as a parallel combination of a capacitor and a
resistor with a switch.\label{fig:DR-circuit}}
\end{figure}

\begin{figure}
\includegraphics[scale=0.7]{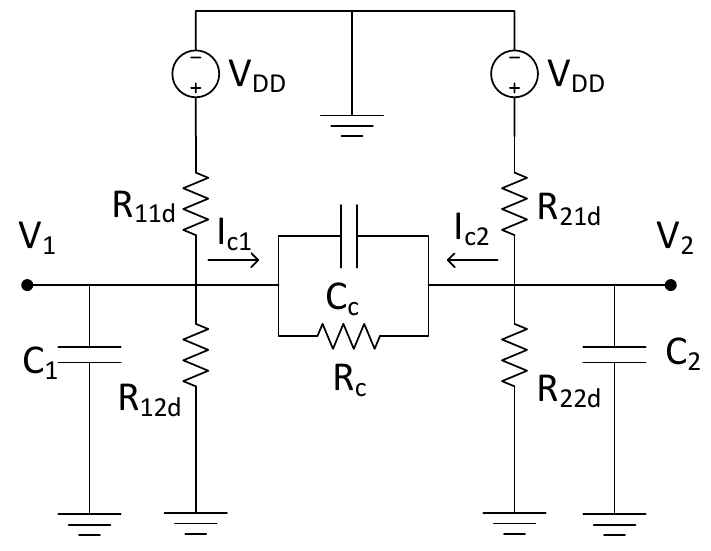}

\protect\caption{Circuit equivalent of coupled D-D oscillators of with an RC circuit
used as the coupling circuit. For D-R oscillators, one state-changing
resistor of each oscillator is replaced by a constant linear resistor\label{fig:coupled-circuit}}
\end{figure}

An electrical circuit representation of the relaxation oscillators
is important to define the form of coupling which is physically realizable.
The basic relaxation oscillator involves repeated charging and discharging
of a capacitor through a resistive path. Switching between charging
and discharging has to be done autonomously by the circuit configuration.
In this paper, we are concerned with the relaxation oscillators built
using state-changing devices. Such state-changing devices are fabricated
using correlated oxide (vanadium dioxide, $VO{}_{2}$) and exhibit
MIT (metal-insulator transition) where the device switches between
a metallic and an insulating state under the application of heat or
an electric field\cite{Kar:2013aa}. Further details about the physical
implementation of these devices are discussed in section \ref{sec:Experimental-verification}.
We will consider two kinds of relaxation oscillator circuits using
such state-changing devices - (a) two state-changing devices in series
(\figref{DD-circuit}). We will refer to this configuration as D-D.
And (b) a state changing device in series with a resistance (\figref{DR-circuit})\cite{Shukla:2014aa}.
This configuration will be referred to as D-R. The D-D configuration
is enticing in its simplicity, both in physical realization and analysis
as will be evident in the following sections. The D-R configuration,
on the other hand, has already been experimentally demonstrated \cite{Shukla:2014aa}
and can be thought of as an extension of the D-D configuration albeit
with more complex dynamics of synchronization. In this paper we will
first study the D-D configuration, using analytical and numerical
techniques; and show through phase models and flow analysis some key
results in the D-R configuration. 

The state transition of the device follow: 
\begin{enumerate}[label=(\alph*),nosep]
\item Only the resistance of the device changes with its state; and the
resistance is linear; 
\item A state transition is triggered by the voltage across the device.
This triggering can be electric field driven or thermally driven,
and can be modeled as an equivalent triggering voltage\cite{Shukla:2014aa}.
When the voltage exceeds a higher threshold $v_{h}$, the state changes
to a metallic (low resistance) state and when the voltage exceeds
a lower threshold $v_{l}$, the state switches back to the insulating
(high resistance) state. The thresholds $v_{h}$ and $v_{l}$ are
not equal, i.e. there is hysteresis in the switching with $v_{l}<v_{h}$,
and 
\item A capacitance is associated with the device that ensures gradual build
up and decaying of the voltage (and hence energy) across the device
$v_{D}$.
\end{enumerate}
The present study of the synchronization dynamics of such coupled
systems, although inspired by the experimental realization of $VO{}_{2}$
based oscillators, is not limited to these oscillators only, but encompasses
a class of similar pairwise-coupled relaxation oscillators as well.
The circuit equivalents of single and coupled relaxation oscillators
are shown in \figref{DD-circuit}, \figref{DR-circuit} and \figref{coupled-circuit}
respectively. The internal resistance of the device $R_{d}$ has two
different values in the two states of the device - $R_{di}$ in the
insulating (high resistance) state and $R_{dm}$ in the metallic (low
resistance) state. $C$ is the internal capacitance of the MIT device
(including any parasitic capacitances) and $R_{S}$ is the series
resistance. We will also assume that $R_{di}\gg R_{dm}$. In the D-D
configuration, the capacitor being charged can be represented as a
single capacitor at the output circuit node. The coupling circuit
is a parallel combination of a capacitor $C_{c}$ and a resistor $R_{c}$.
As shown, the output node of the oscillator is between the device
and the resistance, and the coupling circuit is connected between
these output nodes\cite{Shukla:2014aa}.

\section{Model Development For Isolated \& Coupled Oscillators}

Before investigating the system dynamics, let us establish the system
model and the system of ODEs that define the system. This will allow
us to define the conditions for oscillation as well as the coupling
dynamics. We will first consider D-D configuration and then D-R configuration
as an extension of the D-D configuration. The D-D configuration, owing
to its inherent symmetry renders to easier dynamics and analysis and
provides valuable insights into the system. Such key numerical and
analytical results for this are discussed in the following sections.

\subsection{D-D configuration}

The circuit equivalent for a D-D type relaxation oscillator is shown
in \figref{DD-circuit}. For simplicity, all voltages are normalized
to $v_{dd}$ (including $v_{l}$ and $v_{h}$). We define conductances
$g_{di}=R_{di}^{-1}$, $g_{dm}=R_{dm}^{-1}$ and $g_{c}=R_{C}^{-1}$.
For the conductances, subscript $d$ denotes a state dependent device
conductance and $m/i$ denotes metallic/insulating state respectively.
The subscripts preceding $dm$ or $di$ refer to the corresponding
numbered device as shown in figure. Also, it is assumed that $g_{dm}\gg g_{di}$,
which means that the $g_{di}$ state essentially disconnects the circuit.
This implies that the effective charging happens through $g_{1dm}$
and effective discharging through $g_{2dm}$. The single D-D oscillator
can be described by the following set of piecewise linear differential
equations:
\begin{equation}
cv'=\begin{cases}
(v_{dd}-v)g_{1dm} & charging\\
-v\, g_{2dm} & discharging
\end{cases}
\end{equation}

where $c$ is the lumped capacitance of both devices along with the
parasitics. The equation can be re-written as:
\begin{equation}
cv'=-g(s)v+p(s)\label{eq:eq_s-1}
\end{equation}

where $s$ denotes the conduction state of the device (0 for metallic,
and 1 for insulating) and $g(s)$ and $p(s)$ depend on the device
conduction state $s$ as follows:
\begin{eqnarray}
g(s) & = & \begin{cases}
g_{1dm}, & s=0\\
g_{2dm}, & s=1
\end{cases}\\
p(s) & = & \begin{cases}
g_{1dm}, & s=0\\
0, & s=1
\end{cases}
\end{eqnarray}

When two identical oscillators are coupled in a manner described in
\figref{coupled-circuit}, the system can be described by the following
coupled equations:
\begin{align}
c_{1}v_{1}' & =\begin{cases}
(v_{dd}-v_{1})g_{11dm}-i_{c1} & charging\\
-v_{1}\, g_{12dm}-i_{c1} & discharging
\end{cases}\\
c_{2}v_{2}' & =\begin{cases}
(v_{dd}-v_{2})g_{21dm}-i_{c2} & charging\\
-v_{2}\, g_{22dm}-i_{c2} & discharging
\end{cases}
\end{align}

where $c_{1}$ and $c_{2}$ are the lumped capacitances of the oscillators.
For conductances $g$, the first subscript denotes the oscillator
and the second denotes the device. $i_{c1}=-i_{c2}$ is the coupling
current given by:
\begin{equation}
i_{c1}=(v_{1}'-v_{2}')c_{c}+(v_{1}-v_{2})g_{c}
\end{equation}

When coupled, the system has 4 conduction states $s=s_{1}s_{2}\in\{00,01,10,11\}$
corresponding to the 4 combinations of $s_{1}$ and $s_{2}$. Analogous
to (\ref{eq:eq_s-1}), the coupled system can be described in matrix
form as:
\begin{eqnarray}
c_{c}Fx'(t) & = & -g_{c}A(s)x(t)+P(s)\nonumber \\
x'(t) & = & -\frac{g_{c}}{c_{c}}F^{-1}A(s)\left(x(t)-A^{-1}(s)P(s)\right)\label{eq:main-1}
\end{eqnarray}

where $x(t)=\left(v_{1}(t),v_{2}(t)\right)$ is the state variable
at any time instant $t$. The $2\times2$ matrices $F$ and $A(s)$,
and vector $P(s)$ are given by:
\begin{equation}
F=\left[\begin{array}{cc}
1+\alpha_{1} & -1\\
-1 & 1+\alpha_{2}
\end{array}\right]
\end{equation}
\begin{equation}
\begin{array}{cc}
A(00)=\left[\begin{array}{cc}
-\beta_{11}-1 & 1\\
1 & -\beta_{21}-1
\end{array}\right], & P(00)=\left[\begin{array}{c}
\beta_{11}\\
\beta_{21}
\end{array}\right]\\
A(10)=\left[\begin{array}{cc}
-\beta_{12}-1 & 1\\
1 & -\beta_{21}-1
\end{array}\right], & P(10)=\left[\begin{array}{c}
0\\
\beta_{21}
\end{array}\right]\\
A(01)=\left[\begin{array}{cc}
-\beta_{11}-1 & 1\\
1 & -\beta_{22}-1
\end{array}\right], & P(01)=\left[\begin{array}{c}
\beta_{11}\\
0
\end{array}\right]\\
A(11)=\left[\begin{array}{cc}
-\beta_{12}-1 & 1\\
1 & -\beta_{22}-1
\end{array}\right], & P(11)=0
\end{array}
\end{equation}

Here, $\alpha_{i}=c_{i}/c_{c}$ is the ratio of the combined lumped
capacitance of $i^{th}$ oscillator to the coupling capacitance $c_{c}$,
and $\beta_{ij}=g_{ijdm}/g_{c}$ is the ratio of the metallic state
resistance of $j^{th}$ device of $i^{th}$ oscillator, where $i\in\{1,2\}$
and $j\in\{1,2\}$. $ $The fixed point in a conduction state $s$
is given by $p_{s}=A^{-1}(s)P(s)$ and the matrix determining the
flow (the \emph{flow matrix} or the \emph{velocity matrix}) is given
by $\frac{g_{c}}{c_{c}}F^{-1}A(s)$ as can be seen in \eqref{main-1}.
In section \ref{sec:Symmetric-D-D-coupled} we analyze the steady
state locking and synchronization dynamics of two such identical oscillators
coupled with a parallel resistive and capacitive element as shown
in \figref{coupled-circuit}.

\subsection{D-R configuration}

The equivalent circuit for a D-R type relaxation oscillator is shown
in \figref{DR-circuit}. As in the case of D-D configuration, voltages
are normalized to $v_{dd}$. The conductances involved are $g_{di}=R_{di}^{-1}$,
$g_{dm}=R_{dm}^{-1}$, $g_{s}=R_{s}^{-1}$ and $g_{c}=R_{C}^{-1}$.
Effective charging happens through $g_{dm}$ as in the previous case
but there is an added leakage through $g_{s}$, whereas effective
discharging happens only through $g_{s}$. Following the same methodology
as in the D-D case, the equation for the single D-D oscillator dynamics
can be written as:
\begin{equation}
cv'=\begin{cases}
(v_{dd}-v)g_{dm}-v\, g_{s} & charging\\
-v\, g_{s} & discharging
\end{cases}
\end{equation}

which can be re-written as:
\begin{equation}
cv'=-g(s)v+p(s)\label{eq:eq_s-dr-1}
\end{equation}

where,
\begin{eqnarray}
g(s) & = & \begin{cases}
g_{dm}+g_{s}, & s=0\\
g_{s}, & s=1
\end{cases}\\
p(s) & = & \begin{cases}
g_{dm}, & s=0\\
0, & s=1
\end{cases}
\end{eqnarray}

and $s$ denotes the conduction state of the system as before. In
case of coupled D-R oscillators, arguments similar to the previous
case lead to the same matrix equation as (\ref{eq:main-1}):
\begin{equation}
x'(t)=-\frac{g_{c}}{c_{c}}F^{-1}A(s)\left(x(t)-A^{-1}(s)P(s)\right)
\end{equation}

where matrices $F$ and $P$ remain the same as before but matrix
$A$ changes to the following:
\begin{equation}
F=\left[\begin{array}{cc}
1+\alpha_{1} & -1\\
-1 & 1+\alpha_{2}
\end{array}\right]
\end{equation}
\begin{equation}
\begin{array}{cc}
A(00)=\left[\begin{array}{cc}
-\beta_{1}-\beta_{s1}-1 & 1\\
1 & -\beta_{2}-\beta_{s2}-1
\end{array}\right], & P(00)=\left[\begin{array}{c}
\beta_{1}\\
\beta_{2}
\end{array}\right]\\
A(10)=\left[\begin{array}{cc}
-\beta_{s1}-1 & 1\\
1 & -\beta_{2}-\beta_{s2}-1
\end{array}\right], & P(10)=\left[\begin{array}{c}
0\\
\beta_{2}
\end{array}\right]\\
A(01)=\left[\begin{array}{cc}
-\beta_{1}-\beta_{s1}-1 & 1\\
1 & -\beta_{s2}-1
\end{array}\right], & P(01)=\left[\begin{array}{c}
\beta_{1}\\
0
\end{array}\right]\\
A(11)=\left[\begin{array}{cc}
-\beta_{s1}-1 & 1\\
1 & -\beta_{s2}-1
\end{array}\right], & P(11)=0
\end{array}
\end{equation}

Here $\beta_{i}=g_{idm}/g_{c}$ and $\beta_{si}=g_{si}$.

For all numerical simulations in the rest of the paper, the normalized
values of $v_{l}$ and $v_{h}$ w.r.t $v_{dd}$ are chosen to be 0.2
and 0.8 respectively.

\section{Phase Space, Flows and Oscillation Conditions\label{sec:Phase-Space,-Flows}}

\subsection{Single Oscillators}

A series arrangement of two MIT devices (D-D), or an MIT device and
a resistor (D-R) will oscillate only when certain conditions are met.
In case of two devices in series (D-D), the two devices must be in
opposite conduction states (one metallic and the other insulating)
all the time for oscillations to occur. If the threshold voltages
$v_{l}$ and $v_{h}$ are same for the devices and the following condition
holds
\begin{equation}
v_{l}+v_{h}=V_{DD}\label{eq:vl-vh-condition}
\end{equation}
and at $t=0$ the devices are in different conuduction states, then
any time one device switches, the other will make the opposite transition
as well. The basic mechanism of oscillations is as follows. The device
in metallic state connects the circuit and charges (discharges) the
output capacitor, and the other device in insulating state does not
participate in the dynamics. As the capacitor charges, the voltage
drop across the device in metallic state decreases and crosses the
lower threshold $v_{l}$. At the same instant, the voltage drop across
the other device in insulating state increases and crosses the higher
threshold $v_{h}$ because $v_{D1}+v_{D2}=V_{DD}$. The devices then
switch states and the cycle continues. The devices can be conceived
as a switch which is open in insulating state (ignoring any leakage
in the insulating state) and closed in metallic state (\ref{fig:DD-circuit}).
If $v_{l}$ and $v_{h}$ deviate from (\ref{eq:vl-vh-condition}),
the devices will not switch at the same instant and oscillations will
stop as the system settles to a stable point where both devices are
in same state and the voltage of the output nodes remains at $V_{DD}/2$.
This may require additional startup circuit in the system, which is
trivial to integrate.

\begin{figure}
\includegraphics[scale=0.6]{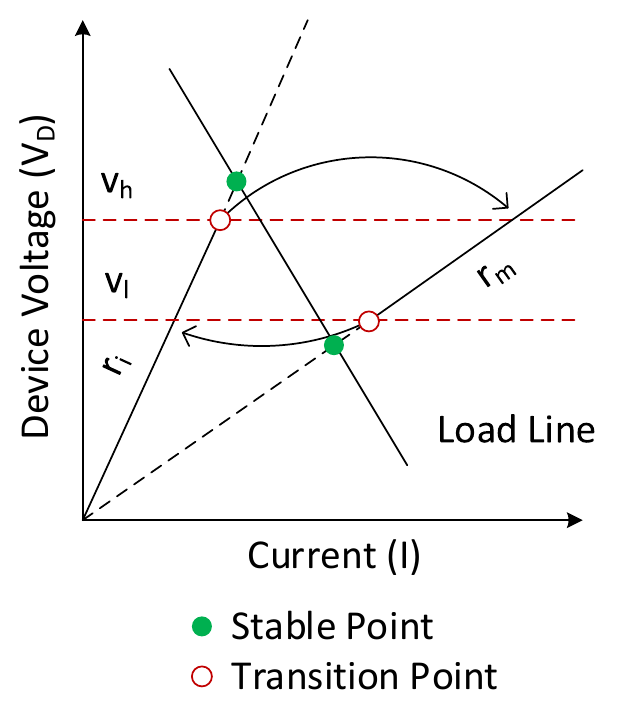}

\protect\caption{Phase space of the device in a single D-R oscillator. Lines with slopes
$r_{i}$ and $r_{m}$ are the regions of operation in insulating and
metallic conduction states respectively. The intersection of these
regions with the load line are the stable points in each region of
operation. The transition points should be encountered before reaching
the stable points for sustained oscillations as shown\label{fig:single-flow}}
\end{figure}

In D-R configuration, another set of conditions have to be met\cite{Hu:1986aa}
which depend on the relative values of the device resistances in the
two states ($R_{dm}$ and $R_{di}$) and the series resistance ($R_{S}$).
These conditions can be described using the phase diagram of the MIT
device \figref{single-flow}. Lines with slopes $r_{i}$ and $r_{m}$
are the regions of operation of the device in insulating and metallic
states respectively. The intersection of these lines with the load
line due to the series resistance gives the stable points of the system
in the two states. For self-sustained oscillations, the stable points
in each state should lie outside the region of operation, i.e. outside
the region defined by horizontal lines passing through the transition
points. This ensures that the system always tries to reach the stable
point in the current state but is always preceded by a transition
to the other state. This moves the system towards the stable point
of the other state (away from the previous stable point) and hence
the system never reaches any stable point and oscillates. This configuration
is robust towards deviation of $v_{l}$ and $v_{h}$ from condition
(\ref{eq:vl-vh-condition}) and as only one device is involved, it
does not require the difficult constraint of simultaneous switching
of devices as was in the D-D case. This reduced requirement of symmetry
is an attractive property of the D-R configuration as initial experiments
have confirmed sustained oscillations in this configuration\cite{Shukla:2014aa}.

We define the region of operation of a device (and hence of an oscillator)
as the region where the device voltage lies between $v_{l}$ and $v_{h}$
(or the output voltage lies between $1-v_{l}$ and $1-v_{h}$). For
the D-D case, the oscillators are expected to remain within the region
of operation all the time. However in the D-R case, the system can
go outside the region of operation in a specific manner as described
later.

\subsection{Coupled Oscillators}

\begin{figure}
\includegraphics[scale=0.6]{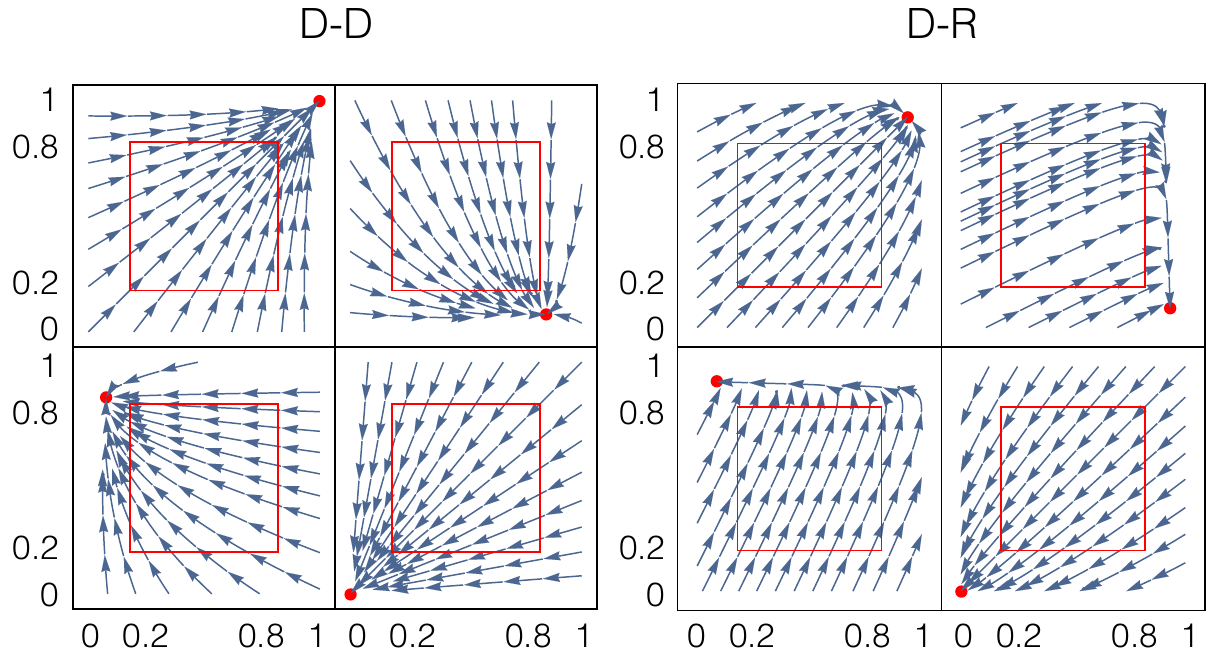}

\protect\caption{The coupled system can be described by 4 different phase spaces for
each state $s=s_{1}s_{2}$. This figure shows the system flows of
the D-D and D-R coupled oscillator system in the 4 regions of operation
along with the fixed points (shown as red dots) in each state. This
figure also represents the simplified case where the flows are monotonic
within the region of operation\label{fig:coupled-flow}}
\end{figure}

For analyzing the coupled circuits, the phase diagram of a coupled
system can be drawn in the $v_{1}\times i_{1}\times v_{2}\times i_{2}$
space as was done in \figref{single-flow}. However, we note that
in a given conductio state of the system, $s=s_{1}s_{2}$, $(v_{1},v_{2})$
can uniquely identify the system, and hence, $v_{1}\times v_{2}$
space is sufficient for a phase diagram. Therefore, we can draw 4
different phase diagrams of the system for each conduction state $s$
(\figref{coupled-flow}) with transitions among them\cite{Saito:1988aa}
(\figref{Transitions}). The transitions occur at the edges when either
$v_{1}$ or $v_{2}$ reach the higher or lower threshold for state
change from metallic to insulating or vice versa. The flows in each
of the 4 conduction states are linear flows and hence have a single
fixed point (\figref{coupled-flow}). The conditions for oscillations
can be described using \figref{coupled-cond}. Analogous to the case
of a single oscillator, these stable points should lie outside the
region of operation (in the shaded region) in a way that the system
always tries to move towards these stable points but should be preceded
by a state transition which occurs when the system reaches the (red)
dashed lines.

\subsubsection{Monotonic Flows and Periodic Orbits}

\begin{figure}
\includegraphics[scale=0.5]{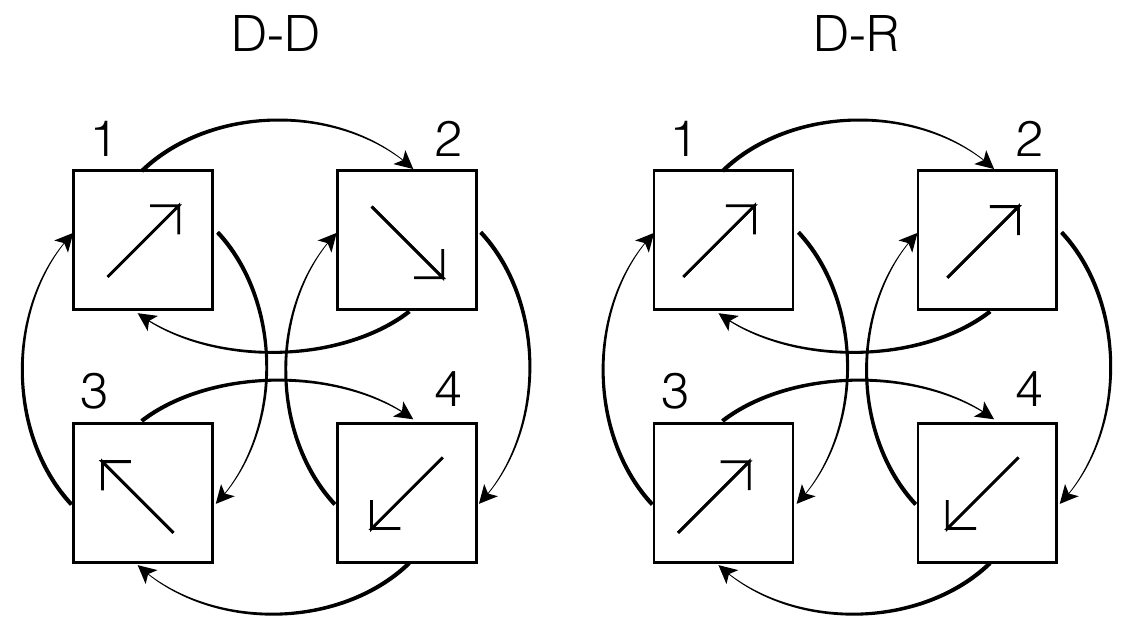}

\protect\caption{Schematic representation showing the monotonic flow directions in
the regions of operation in the simplified model. The monotonicity
condition is sufficient for existence of a steady state periodic orbit
in the D-D case. Transitions are shown among the 4 states 1(MM), 2(IM),
3(MI) and 4(II) of the coupled system when the system reaches any
edge, i.e. the voltage of any oscillator reaches a phase change threshold
of its MIT device\label{fig:Transitions}}
\end{figure}

\begin{figure}
\includegraphics[scale=0.7]{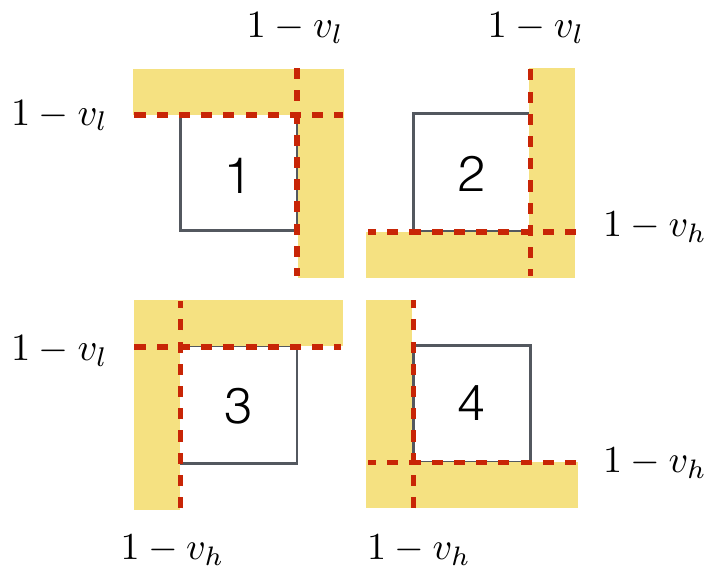}

\protect\caption{The stable points of both D-D and D-R coupled oscillator system should
lie in the yellow shaded region for the system to oscillate. The system
undergoes a transition to another state when the system hits the red
dashed lines\label{fig:coupled-cond}}
\end{figure}

\begin{figure}
\includegraphics[scale=0.8]{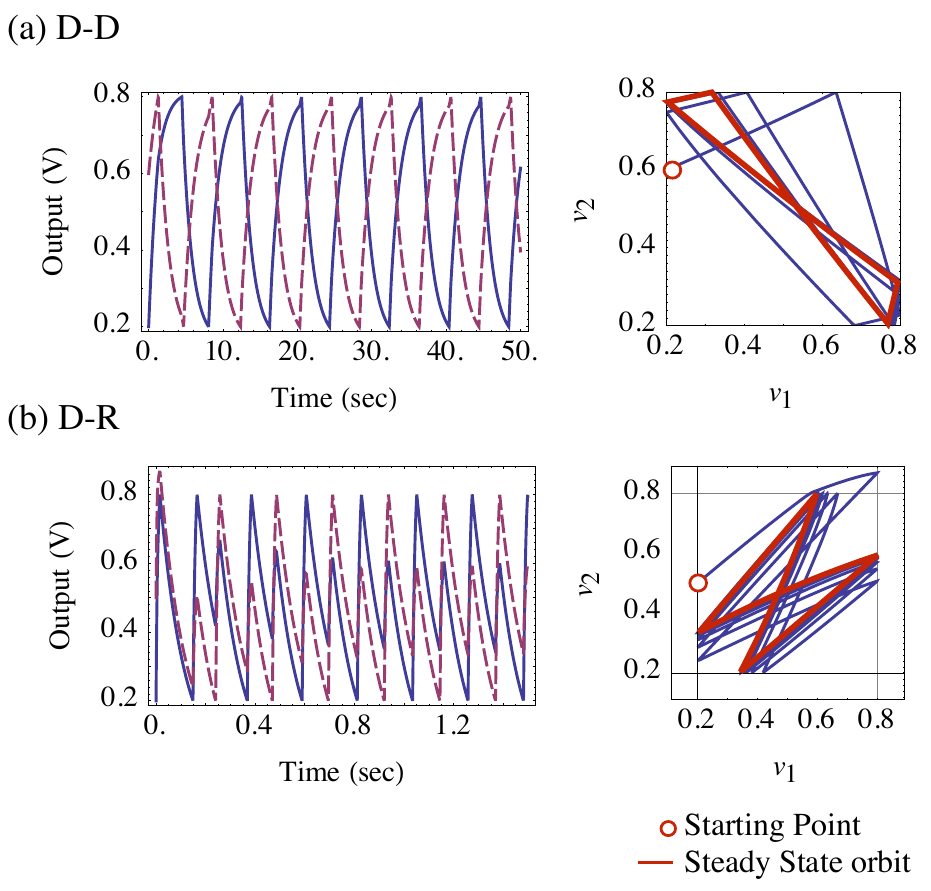}\protect\caption{Simulation waveforms with time (left) and the system trajectory in
phase space (right) of a system of coupled oscillators of type (a)
D-D and (b) D-R. The steady state periodic orbit is shown in red.
The butterfly shaped steady state trajectory corresponds to waveforms
similar to anti-phase locking. The solid and dashed lines represent
the two oscillators.\label{fig:Simulation-waveforms}}
\end{figure}

The conditions of \figref{coupled-cond} are general enough to hold
for both D-D and D-R configurations and they ensure that the system
does not settle down to a stable point and voltages across oscillators
repeatedly increase and decrease. However, these conditions do not
ensure the existence of a stable orbit which can give periodic oscillations.
To ensure existence of a stable periodic orbit, we consider additional
conditions for the systems. For D-D configuration, we consider systems
where the flows in the states are monotonic, i.e. $v_{1}$ and $v_{2}$
are either constantly increasing or constantly decreasing in the region
of operation of any conduction state. \Figref{Transitions} show these
monotonic directions with the state transitions for D-D coupled oscillator
configurations. It is proved later that for two identical coupled
D-D oscillators, this condition of monotonicity of the flows is \emph{sufficient}
for existence of a stable orbit and hence for periodic oscillations.
For D-R coupled oscillators, we consider systems where either the
direction of flows are strictly monotonic as shown in \figref{Transitions}
or are non-monotonic in a very specific way as discussed in \ref{sec:D-R-coupled-oscillator}
(see \figref{dr-symmetry-reduced}). In this case, periodic oscillations
can be ensured for certain conditions as described in section \ref{sec:D-R-coupled-oscillator}.
It should be noted here that in the D-R case, the system can also
go outside the region of operation as seen in \figref{Simulation-waveforms},
but if the fixed points lie in the above mentioned shaded regions,
the system will always oscillate. \Figref{Simulation-waveforms} shows
typical time-domain waveforms and corresponding phase-space trajectories
for the coupled oscillators of the D-D and D-R types.

\section{Symmetric D-D coupled oscillator dynamics\label{sec:Symmetric-D-D-coupled}}

\begin{figure}
\includegraphics[scale=0.6]{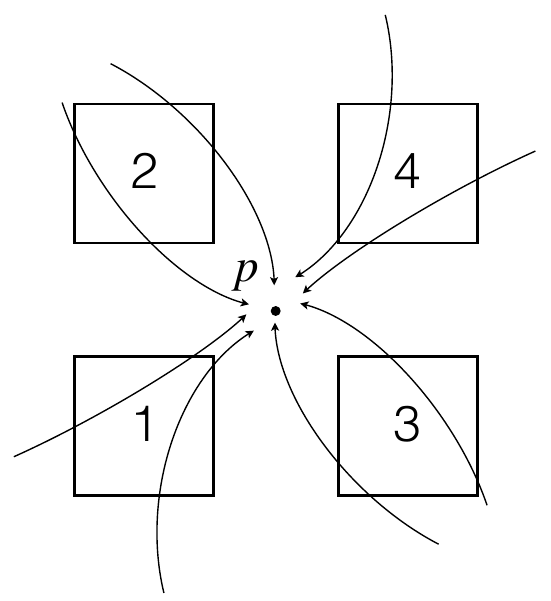}

\protect\caption{Combined phase space showing 4 regions of operation of the four different
conduction states such that all states share a single fixed point
$p$. This is possible as the flow matrices in all the four states
are equal and, hence, all state spaces can be represented in a single
space with a single flow but occupying different regions\label{fig:dd-combined-phase-space}}
\end{figure}

Let us first investigate the case when the D-D oscillators are identical
and their effective charging and discharging rates are equal, i.e.
$\beta_{11}=\beta_{21}=\beta_{12}=\beta_{22}=\beta$ and $\alpha_{1}=\alpha_{2}$.
This corresponds to a well designed and ideal oscillator system where
the pull-up and pull-down device resistances have been matched to
create equal charging and discharging rates. In such a scenario the
velocity matrices in the four conduction states $\frac{g_{c}}{c_{c}}F^{-1}A(s)$
become equal. As such, the state spaces in the four conduction states
can be represented in a common state space with the system flow described
by the common velocity matrix and a single fixed point. However, in
this common state space, the regions of operation in the four conduction
states will be four distinct regions. The position of these regions
for a conduction state would depend on the position of its respective
fixed points in the original state space. Such a combined phase space
is shown in \figref{dd-combined-phase-space}.

The symmetry of the system is apparent in the flow as well. The eigen
values $\lambda_{1},\lambda_{2}$ and eigen vectors $e_{1},e_{2}$
of the velocity matrix $\frac{g_{c}}{c_{c}}F^{-1}A$ of the symmetric
system are 
\begin{equation}
\lambda_{1}=-\frac{g_{c}}{c_{c}}\left(\frac{\beta}{\alpha}\right),\,\lambda_{2}=-\frac{g_{c}}{c_{c}}\left(\frac{\beta+2}{\alpha+2}\right)
\end{equation}

\begin{equation}
e_{1}=\left[\begin{array}{c}
1\\
1
\end{array}\right],\, e_{2}=\left[\begin{array}{c}
-1\\
1
\end{array}\right]
\end{equation}

\begin{figure}
\includegraphics[scale=0.6]{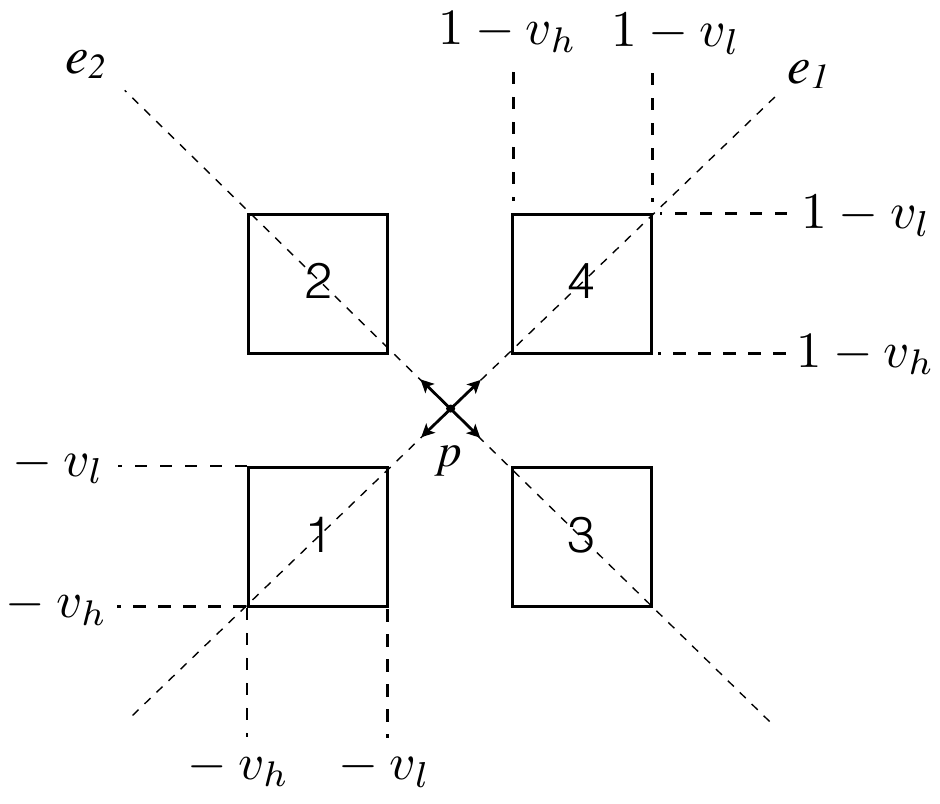}

\protect\caption{In the combined phase space, the flow of the coupled system is a mirror
image of itself about its eigen vector directions $e_{1}$ and $e_{2}$
as the eigen values are real and negative. This symmetry of the flows
can be reduced and the state space of the system can be described
by considering just one-fourth of this space as shown in \figref{dd-sym-symmetry-reduced}\label{fig:dd-sym-symmetry}}
\end{figure}

\begin{figure}
\includegraphics[scale=0.7]{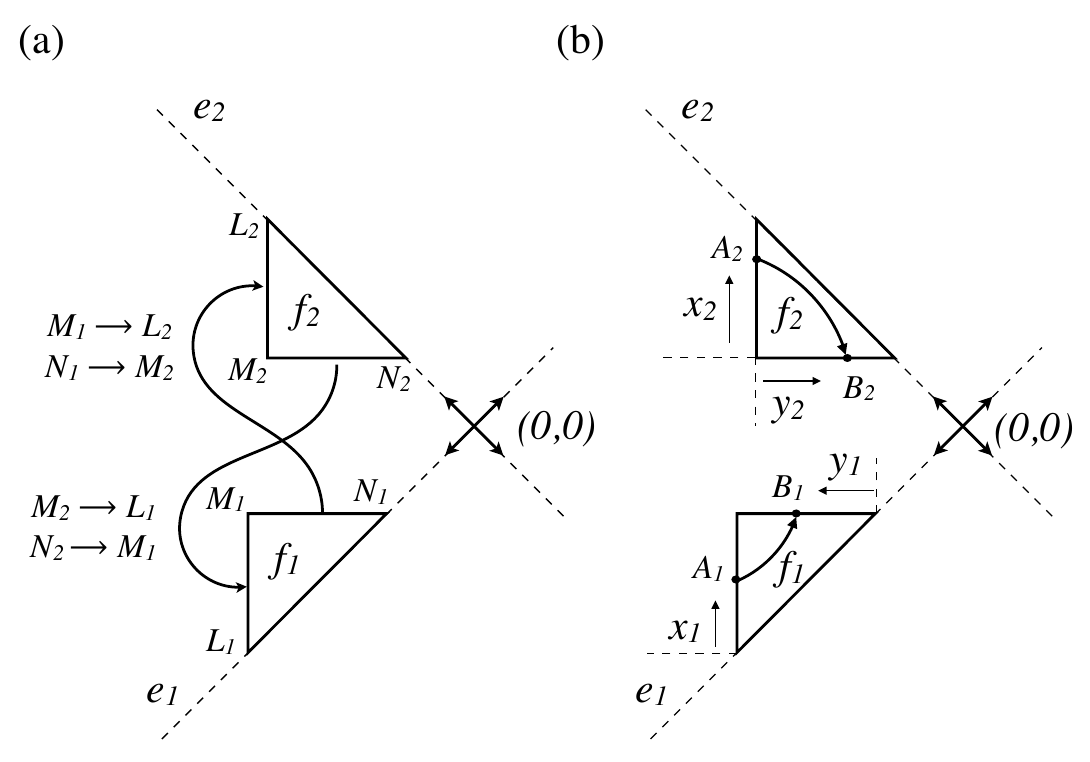}

\protect\caption{(a) Symmetry reduced space (fundamental domain) of the coupled system
after reducing the symmetries shown in \figref{dd-sym-symmetry}.
$f_{1}$ is the mapping from the left edge of state 1 to its top edge
and $f_{2}$ is the mapping from left edge of state 2 to its bottom
edge. (b) Definition of $x_{1},\, x_{2},\, y_{1}$ and $y_{2}$ on
the edges of the states in the symmetry reduced space\label{fig:dd-sym-symmetry-reduced}}
\end{figure}

Real negative eigen values imply that the flow of the system is symmetric
about both the eigen vector directions (i.e. a mirror image of itself
about the eigen directions) as shown in \figref{dd-sym-symmetry}.
The stable fixed points in the conduction states 1(00), 2(01), 3(10)
and 4(11) are $p_{1}=(1,1)$, $p_{2}=\left(1-\frac{1}{2+\beta},\frac{1}{2+\beta}\right)$,
$p_{3}=\left(\frac{1}{2+\beta},1-\frac{1}{2+\beta}\right)$ and $p_{4}=(0,0)$
respectively. Hence, the line along the eigen vector $e_{1}$ is the
diagonal for both conduction states 1 and 4. Under the assumption
that the $v_{dd}$ normalized thresholds $v_{l}$ and $v_{h}$ are
symmetric i.e. $v_{l}=1-v_{h}$, the line along $e_{2}$ also becomes
the diagonal for states 2 and 3. This is because the fixed points
of conduction states 2 and 3 - $p_{2}$ and $p_{3}$ lie on $x+y=1$
line in their original state spaces which is same as the eigen direction
$e_{2}$. It should now be noted that the transitions between the
conduction states, the regions of operation and the flow, all have
the same common discrete symmetry - mirroring about $e_{1}$ and $e_{2}$.
We can do a symmetry reduction at this point and the system can be
completely described by just two states and two transitions (\figref{dd-sym-symmetry-reduced}a).

\begin{figure}
\includegraphics[scale=0.8]{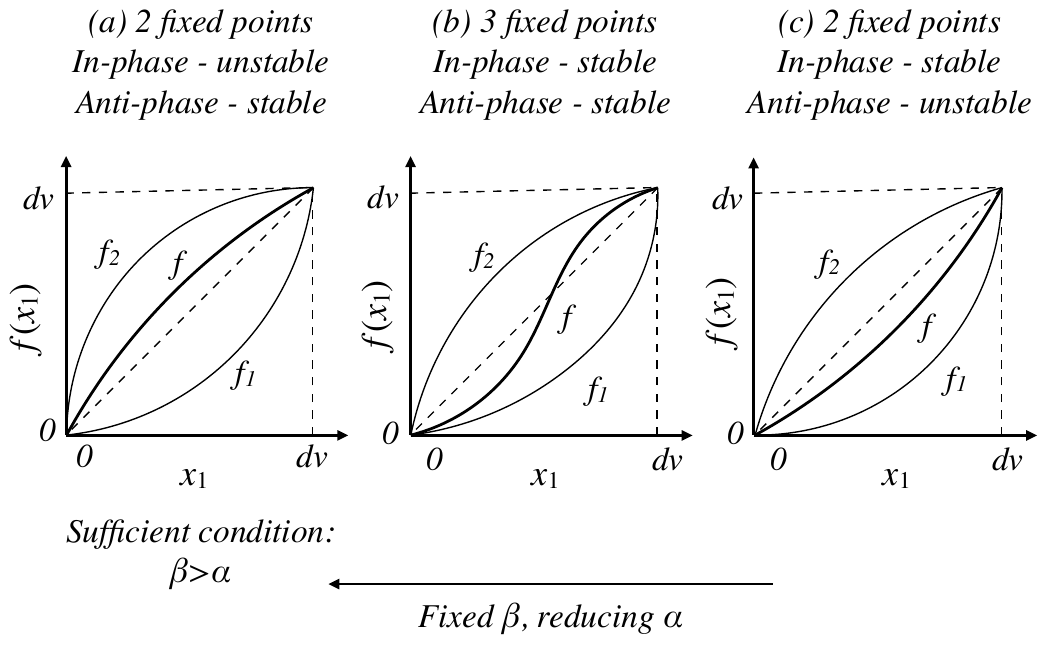}

\protect\caption{Representative plot of mappings $f_{1}$, $f_{2}$ and their composition
$f=f_{1}\circ f_{2}$ with fixed $\beta$ and varying $\alpha$. Here
$dv=v_{h}-v_{l}$. $\beta>\alpha$ is a sufficient condition for a
concave $f$ and hence stable anti-phase locking. As $\alpha$ increases,
the curve for $f$ transitions into a s-shaped curve with both in-phase
and anti-phase lockings stable, and then finally to a convex curve
with stable in-phase locking\label{fig:dd-sym-return-maps}}
\end{figure}

\begin{figure}
\includegraphics[scale=0.8]{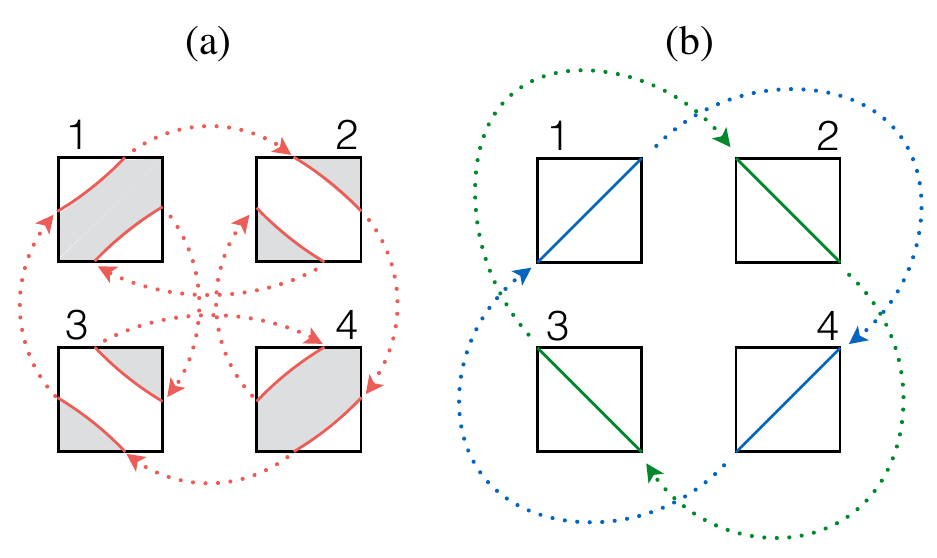}

\protect\caption{The trajectories (which are periodic orbits) corresponding to the
fixed points in the return maps of \figref{dd-sym-return-maps}b.
(a) The unstable fixed point of \figref{dd-sym-return-maps}b corresponds
to two periodic orbits in the unreduced space as shown in red. (b)
The fixed point at 0 corresponds to a single periodic orbit shown
in blue and the fixed point at $dv$ corresponds to the green periodic
orbit. When the initial state of the system lies in the gray region
(shown in (a)), the system settles down to an in-phase locking state,
and otherwise to an anti-phase locking state\label{fig:dd-sym-orbits}}
\end{figure}

To study the steady state periodic orbits of this system, we calculate
the return map on the left edge of state 1 in \figref{dd-sym-symmetry-reduced}a
which is $f=f_{1}\circ f_{2}$. In this case, any periodic orbit in
the symmetry reduced space will correspond to at least one periodic
orbit in the complete space (see \figref{dd-sym-orbits}). Also, if
no fixed point exist in the symmetry reduced space, then there is
definitely no periodic orbit in the complete space. The coordinate
measurements on the edges are defined as shown in \figref{dd-sym-symmetry-reduced}b.
$f_{1}:x_{1}\rightarrow y_{1}$ is the mapping from the left edge
of state 1 to its top edge and $f_{2}:x_{2}\rightarrow y_{2}$ is
the mapping from left edge of state 2 to its bottom edge. $x_{1},x_{2},x_{3}$
and $x_{4}$ are defined on their respective edges as shown in \figref{dd-sym-symmetry-reduced}b.
As both the eigen values $\lambda_{1}$ and $\lambda_{2}$ are real
and negative, $f_{1}(x)$ will lie above $x=y$ line and $f_{2}(x)$
will lie below it. A representative plot of $f_{1}$, $f_{2}$ and
$f=f_{1}\circ f_{2}$ (i.e. the return map $f:x_{1}\rightarrow y_{2}$)
is shown in \figref{dd-sym-return-maps} where $dv=v_{h}-v_{l}$.
The composition $f=f_{1}\circ f_{2}$ lies above $x=y$ if $f_{2}$
is more curved than $f_{1}$ and vice versa. As the return map is
always increasing, only the first return map needs to be considered
for finding fixed points and the higher return maps don't add new
fixed points. When the coupling is more capacitive, the composition
function tends to be concave as shown in \figref{dd-sym-return-maps}a.
\emph{Proposition} 1 gives a mathematical form to this notion.

If the system moves from any arbitrary point on the flow, say $(x_{a},y_{a})$
to another point, $(x_{b},y_{b})$ in time $t$ then the following
implicit equation can be written: 
\begin{eqnarray}
\left(\frac{x_{a}+y_{a}}{x_{b}+y_{b}}\right)^{\frac{1}{\lambda_{1}}} & = & \left(\frac{x_{a}-y_{a}}{x_{b}-y_{b}}\right)^{\frac{1}{\lambda_{2}}}\label{eq:flow-equation}
\end{eqnarray}

In state 1, $(x_{a},y_{a})$ lies on the left edge and $(x_{b},y_{b})$
lies on the top edge. To define $f_{1}:x_{1}\rightarrow y_{1}$ we
substitute $(x_{a},y_{a})=(-v_{h},-v_{h}+x_{1})$ in \eqref{flow-equation}
and obtain an implicit equation for $f_{1}$ as: 
\begin{equation}
\left(\frac{2v_{h}-x_{1}}{2v_{l}+y_{1}}\right)=\left(\frac{x_{1}}{y_{1}}\right)^{\frac{\alpha+2}{\beta+2}\frac{\beta}{\alpha}}\label{eq:flow-eq1}
\end{equation}

Similarly, an implicit equation for $f_{2}:x_{2}\rightarrow y_{2}$
can be written as: 
\begin{equation}
\left(\frac{k_{\beta}+x_{2}}{k_{\beta}-y_{2}}\right)=\left(\frac{dv-x_{2}}{dv-y_{2}}\right)^{\frac{\beta+2}{\alpha+2}\frac{\alpha}{\beta}}\label{eq:flow-eq2}
\end{equation}

where $k_{\beta}=\frac{\beta}{\beta+2}$.

Equations \ref{eq:flow-eq1} and \ref{eq:flow-eq2} can be solved
numerically to obtain the steady state orbits of the system. \\

\emph{Proposition 1} : \emph{Sufficient condition for out-of-phase
locking:} For $\beta>\alpha>\frac{2dv}{1-dv}=\frac{dv}{v_{l}}$, i.e.,
$\frac{g_{dm}}{g_{c}}>\frac{c}{c_{c}}>\frac{2dv}{1-dv}=\frac{dv}{v_{l}}$
the coupled symmetric and identical system has only two steady state
locking orbits - in-phase and out-of-phase. Further, the in-phase
locking is unstable and the out-of-phase locking is stable.\\

\emph{Proof} : The proof can be divided in two steps - (a) There are
only two fixed points of $f$ - at 0 and at $dv$, and (b) $f'(0)>1$
and $f'(dv)<1$ which implies that the in-phase locking is unstable
and anti-phase locking is stable.

The first part is proved as follows.

As $\lambda_{1}$ and $\lambda_{2}$ are negative, $x_{1}>y_{1}$
and $dv-x_{2}>dv-y_{2}$. And as $\beta>\alpha>\frac{2dv}{1-dv}$,
$\frac{\alpha+2}{\beta+2}\frac{\beta}{\alpha}>1$ and $\frac{\beta+2}{\alpha+2}\frac{\alpha}{\beta}<1$.
Also $\beta>\frac{2dv}{1-dv}$ implies $k_{\beta}>dv>y_{2}$. This
gives us the following inequalities: 
\begin{eqnarray}
\left(\frac{2v_{h}-x_{1}}{2v_{l}+y_{1}}\right) & \geq & \left(\frac{x_{1}}{y_{1}}\right)\label{eq:ineqf1}\\
and\,\left(\frac{k_{\beta}+x_{2}}{k_{\beta}-y_{2}}\right) & \leq & \left(\frac{dv-x_{2}}{dv-y_{2}}\right)\label{eq:ineqf2}
\end{eqnarray}

where the equality holds at the end points i.e. at $x_{1}=0$ and
$x_{1}=dv$ for (\ref{eq:ineqf1}) and at $x_{2}=0$ and $x_{2}=dv$
for (\ref{eq:ineqf2}). At any fixed point for the return map $f$,
$x_{1}=y_{2}$ and $y_{1}=x_{2}$ and equations (\ref{eq:ineqf1})
and (\ref{eq:ineqf2}) should be consistent with these fixed point
equations. Substituting $x_{1}=y_{2}$ and $y_{1}=x_{2}$ in (\ref{eq:ineqf1})
and (\ref{eq:ineqf2}) we get: 
\begin{eqnarray}
dv-((dv-y_{1})+y_{2})+\frac{2(dv-y_{1})y_{2}}{dv+k_{\beta}} & \geq & 0\label{eq:final-ineq1}\\
dv-((dv-y_{1})+y_{2})+\frac{(dv-y_{1})y_{2}}{v_{h}} & \leq & 0\label{eq:final-ineq2}
\end{eqnarray}

These equations are consistent only when 
\begin{equation}
(dv-y_{1})y_{2}\frac{2}{dv+k_{\beta}}\geq(dv-y_{1})y_{2}\frac{1}{v_{h}}
\end{equation}

which in turn can be true only at the end points, i.e. $y_{1}=0$
or $y_{1}=dv$, because $k_{\beta}<1$. It can be confirmed that this
is indeed the case by inspection of \figref{dd-sym-symmetry-reduced}.

The second part of the proof is proved by calculating $f'(0)=f_{1}'(0)\cdot f_{2}'(0)$.
$f_{1}'(0)$ and $f_{2}'(0)$ are calculated from (\ref{eq:flow-eq1})
and (\ref{eq:flow-eq2}) as: 
\begin{eqnarray}
f_{1}'(0) & = & \left(\frac{v_{l}}{v_{h}}\right)^{q}\\
f_{2}'(0) & = & \frac{k_{\alpha}+dv}{k_{\alpha}-dv}=\frac{k_{\alpha}+v_{h}-v_{l}}{k_{\alpha}-v_{h}+v_{l}}>\frac{v_{h}}{v_{l}}
\end{eqnarray}

where $q=\frac{\beta+2}{\alpha+2}\frac{\alpha}{\beta}<1$ and $k_{\alpha}=\frac{\alpha}{\alpha+2}$.
Also $\alpha>\frac{2dv}{1-dv}$ implies $k_{\alpha}>dv$. Hence 
\begin{equation}
f'(0)=f_{1}'(0)\cdot f_{2}'(0)>\left(\frac{v_{h}}{v_{l}}\right)^{1-q}>1
\end{equation}

And as $f$ has no other fixed points between 0 and $dv$ and $f$
is continuous, $f'(dv)<1$. Hence, proved. \\

It should be noted that this condition is not a strict bound but rather
provides key design insights when a particular form of coupling (anti-phase)
is sought\cite{Datta:2014aa}.

\subsection{Capacitive, Resistive Coupling and Bistable Orbits}

\begin{figure}
\includegraphics[scale=0.8]{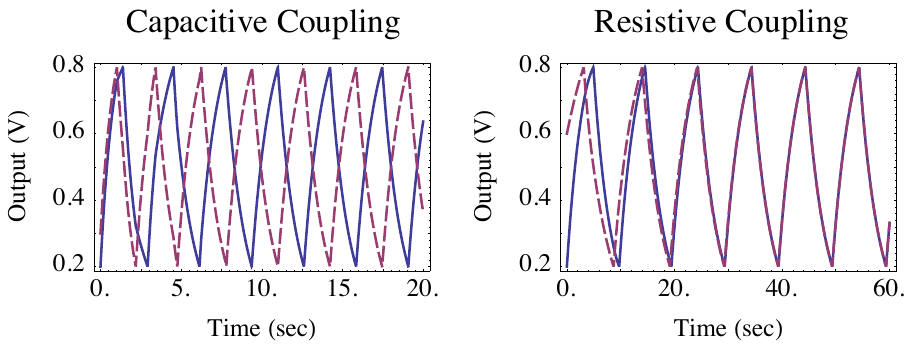}

\protect\caption{Capacitive coupling leads to anti-phase locking and resistive coupling
leads to in-phase locking in case of symmetric D-D coupled. The solid
and dashed lines represent the two oscillators. \label{fig:dd-sym-cap-res}}
\end{figure}

\begin{figure}
\includegraphics[scale=0.8]{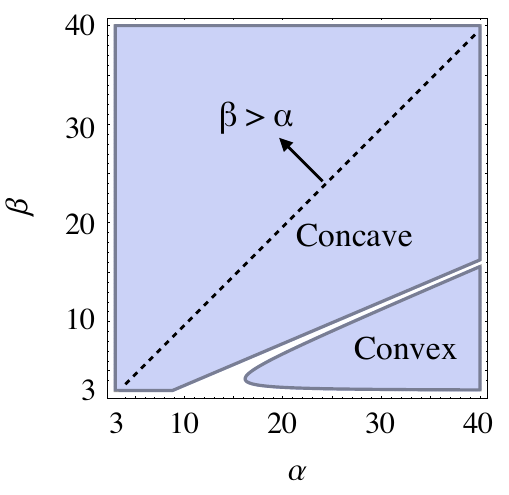}

\protect\caption{Return map type for the symmetric D-D case in the parametric space,
$\beta\times\alpha$ for $v_{l}=0.2$ and $v_{h}=0.8$. We can clearly
see that for $\beta>\alpha$ the return map is concave and anti-phase
locking is stable. Also when the coupling is more resistive, the return
map becomes convex with stable in-phase locking. The region between
concave and convex return map is the region with S-shaped return map
with both stable in-phase and stable anti-phase locking\label{fig:dd-sym-beta-alpha}}
\end{figure}

The two extreme cases of purely resistive and purely capacitive coupling
are of interest. In case of coupling using only a capacitor, the symmetric
and identical coupled system always has a stable anti-phase and an
unstable in-phase locking. This is because in case of purely capacitive
coupling, $\beta\rightarrow\infty$ and so $\beta>\alpha$ for all
finite $\alpha$. Even in practical cases where some parasitic resistance
is included in parallel with the coupling capacitor \cite{Shukla:2014aa},
$\beta$ is typically much larger than $\alpha$. Such anti-phase
locking matches well with recent experimental findings of capacitively
MIT coupled oscillators as discussed in \cite{Datta:2014aa}. In case
of coupling using only a resistor, the symmetric and identical coupled
system will have a stable in-phase and an unstable anti-phase orbit,
as can be predicted from \figref{dd-sym-beta-alpha} for $\alpha\rightarrow\infty$.
Time domain simulations of the coupled systems with purely capacitive
and purely resistive coupling are shown in \figref{dd-sym-cap-res}.
The parameter values for capacitive coupling are $\alpha=5$ and $g_{dm}=g_{s}=6c_{c}$
and those for resistive coupling are $c=13g_{c}$ and $\beta=3.6$. 

\begin{figure}
\includegraphics[scale=0.8]{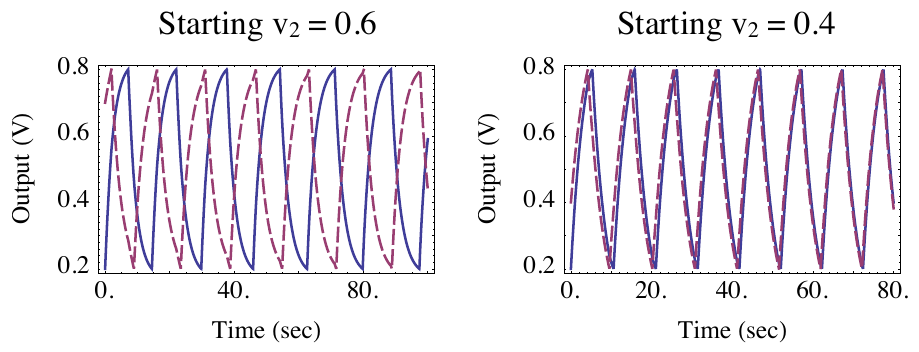}

\protect\caption{Simulation waveforms showing the dependence of final locking to the
initial state of the system in the intermediate case of \figref{dd-sym-return-maps}b
when the return map is S-shaped. The solid and dashed lines represent
the two oscillators. Initial $v_{1}=0.2V$ in both cases, but the
system locks in-phase when initial $v_{2}=0.4V$, and anti-phase when
initial $v_{2}=0.6V$. With reference to \figref{dd-sym-orbits},
the intial point $(0.2,0.4)$ lies in the gray region and the point
$(0.2,0.6)$ lies outside the gray region in conduction state 1\label{fig:dd-sym-bistable-waveforms}}
\end{figure}

\Figref{dd-sym-return-maps} (b and c) show cases when $\beta<\alpha$.
In the intermediate case when the return map transitions from concave
to convex, the system goes through a state where both in-phase and
anti-phase locking are stable with one unstable fixed point in between
(\figref{dd-sym-return-maps}b). In \figref{dd-sym-beta-alpha} the
two regions for concave and convex return map can be clearly seen.
They are separated by a thin region which represents the case of bistability.
\Figref{dd-sym-bistable-waveforms} shows the time domain simulation
waveforms of oscillator outputs for $\beta=3.6$ and $\alpha=13.1$.
We note that the initial voltage of the first oscillator is 0.2V and
depending on the initial voltage of the second oscillator, the system
can either lock in phase or out of phase. These design parameters
correspond to a bistable system of the kind shown in \figref{dd-sym-return-maps}b,
and hence the final steady state locking is in-phase or out-of-phase
depending on the initial phase of the system. When the initial phase
(or output voltage) of oscillators are close to each other (represented
by gray region in \figref{dd-sym-orbits}a) the system locks in-phase,
and when they are far the system locks out-of-phase for the same circuit
parameters. 

\section{Asymmetric D-D coupled oscillator dynamics\label{sec:Asymmetric-D-D-coupled}}

\begin{figure}
\includegraphics[scale=0.7]{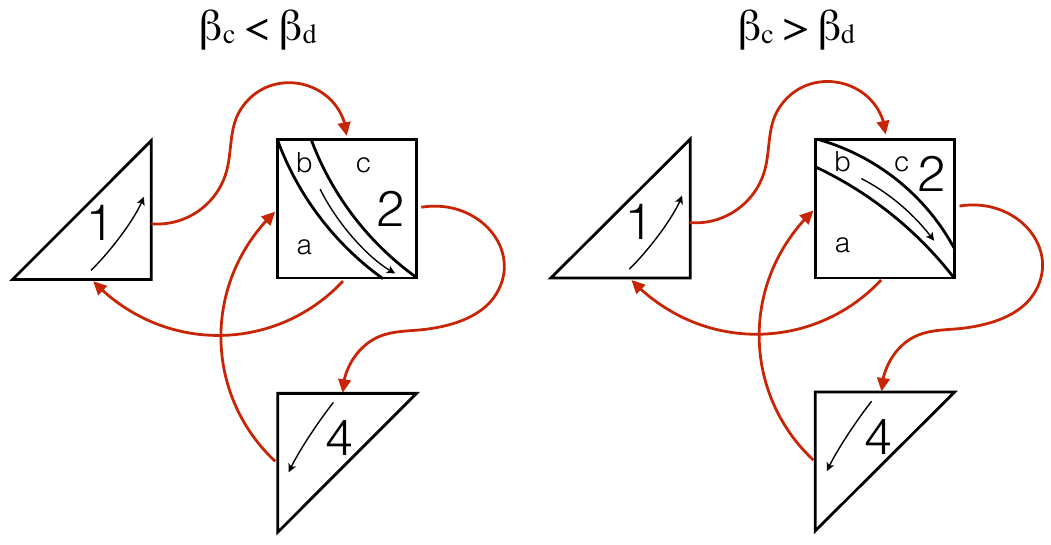}

\protect\caption{Symmetry reduced space in the asymmetric D-D configuration with $\beta_{c}>\beta_{d}$
(left) and $\beta_{c}<\beta_{d}$ (right). Such configuration will
have only a single symmetry. The flow matrices in the four conduction
states are not equal and hence states cannot be represented in a single
combined state space with a single fixed point as was done in the
symmetric D-D case $(\beta_{c}=\beta_{d})$\label{fig:dd-asym-symmetry-reduced}}
\end{figure}

\begin{figure}
\includegraphics[scale=0.6]{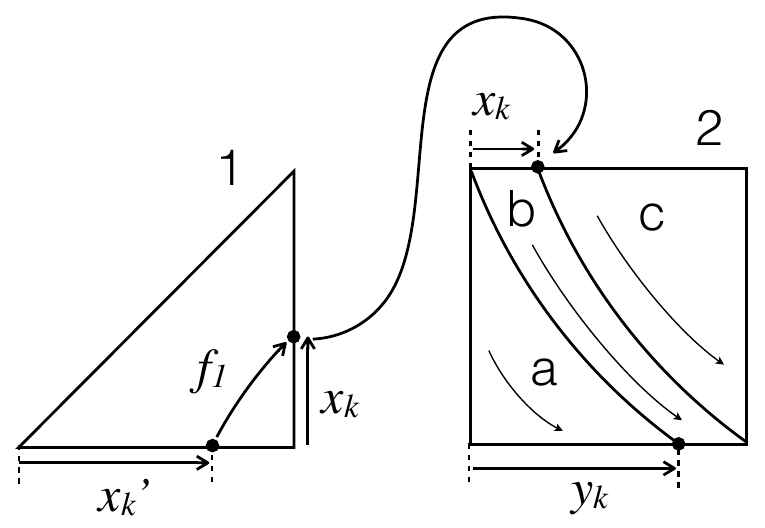}

\protect\caption{Diagram of symmetry reduced state spaces for conduction states 1 and
2 in the asymmetric D-D configuration. In conduction state 2, the
top-left corner does not map to the bottom-right corner as was the
case in the symmetric D-D case. The width of this middle region where
the flow maps the top edge to the bottom edge is defined using $x_{k}$
and $y_{k}$\label{fig:dd-asym-xk}}
\end{figure}

Let us now investigate the case of D-D oscillator dynamics where the
two oscillators are identical but the pull-up and pull-down devices
are non-identical thereby giving rise to asymmetric charging and discharging
rates. As the the oscillators are identical, $\beta_{11}=\beta_{21}=\beta_{c}$
and $\beta_{12}=\beta_{22}=\beta_{d}$ where subscripts $c$ and $d$
stand for charging and discharging. The symmetry of the system (due
to the identical oscillators) can be seen in the flows of the states.
Flows of conduction states 1(00) and 4(11) are mirror images about
the diagonal $x=y$ and the flow in conduction state 2(10) is equivalent
to the flow in state 3(01) with axes x and y interchanged. This symmetry
is also shown in the transitions between states. The system can be
expressed after reducing the symmetry as in \figref{dd-asym-symmetry-reduced}.
For $\beta_{c}<\beta_{d}$, two kinds of cycles are possible in the
regions $1\rightarrow2b\rightarrow1$ and $1\rightarrow2c\rightarrow4\rightarrow2a\rightarrow1$.
To find the fixed points of the system, we draw the return map with
the bottom edge of state 1 as the Poincare section. Because it is
a symmetry reduced space, we consider the first return map for trajectories
of the type $1\rightarrow2c\rightarrow4\rightarrow2a\rightarrow1$
and the second return map for trajectories of the type $1\rightarrow2b\rightarrow1$.
Let $f_{1}(x_{k}')=x_{k}$ as shown in \figref{dd-asym-xk} where
$f_{1}$ is the mapping from bottom edge of conduction state 1 to
its right edge, and $f_{2a},f_{2b}$ and $f_{2c}$ are the mappings
between edges in conduction state 2 as shown. Then the return map
$f$ is given by:
\begin{equation}
f(x)=\begin{cases}
f_{1}\circ f_{2c}\circ f_{4}\circ f_{2a}(x), & 0\leq x<x_{k}'\\
f_{1}\circ f_{2b}\circ f_{1}\circ f_{2b}(x), & x_{k}'\leq x<dv
\end{cases}\label{eq:asym-return-map}
\end{equation}

\emph{\\Proposition 2} : \emph{Existence of stable periodic orbit
in asymmetric D-D coupled oscillator system:} If $f$ is the return
map for the D-D asymmetric coupled oscillator on the bottom edge of
state 1 then the following are true:
\begin{enumerate}[label=(\alph*),nosep]
\item $f$ is continuous
\item $f'(0)>1$ for $\beta_{c}>\alpha$ and $\beta_{d}>\alpha$
\item $f$ has one fixed point at 0 and at least one in the interval $x_{k}'<x<dv$
at, say, $x_{f}$
\item Either the fixed point at $x_{f}$ is stable, or there exists a stable
fixed point at $x_{f}'$ where $0\leq x_{f}'<x_{f}$\\
\end{enumerate}
\emph{Proof }: (a) The return map is separately continuous in intervals
$[0,x_{k}')$ and $(x_{k}',dv]$ as it is a composition of mappings
of continuous flows. The continuity of $f$ at $x_{k}$ can be established
by considering two points close to $x_{k}'$ on either side. From
(\ref{eq:asym-return-map}) we can see that $f(x_{k+}')=f(x_{k-}')=y_{k}$,
and hence $f$ is continuous at $x_{k}$ .

(b) It can be proved by similar procedure as adopted before in \emph{Proposition
1} that $f'(0)=f_{1}'(0)\cdot f_{2c}'(dv)\cdot f_{4}(0)\cdot f_{2a}'(0)>1$
for $\beta_{c}>\alpha$ and $\beta_{d}>\alpha$. 

(c) The fixed point at 0 can be seen clearly in the flow diagram.
In interval $x_{k}'<x<dv$, the fixed points of first return $f_{1}\circ f_{2b}$
will also be the fixed points of second return (which is $f$), but
not the other way around. Now $f_{1}\circ f_{2b}(x_{k}')=dv$ and
$f_{1}\circ f_{2b}(dv)=y_{k}$. As $f_{1}\circ f_{2b}$ is continuous,
and hence decreasing, in the interval $x_{k}'<x<dv$, there exists
a fixed point for $f_{1}\circ f_{2b}$, and hence for $f$, in the
interval $x_{k}'<x<dv$.

(d) As $f$ is continuous and has fixed points at $0$ and $x_{f}$,
one of these two should be stable if there is no other fixed point
in between $0$ and $x_{f}$. If they both are unstable, then a stable
fixed point exists in the interval $(0,x_{f})$. Hence proved.\\

\begin{figure}
\includegraphics[scale=0.6]{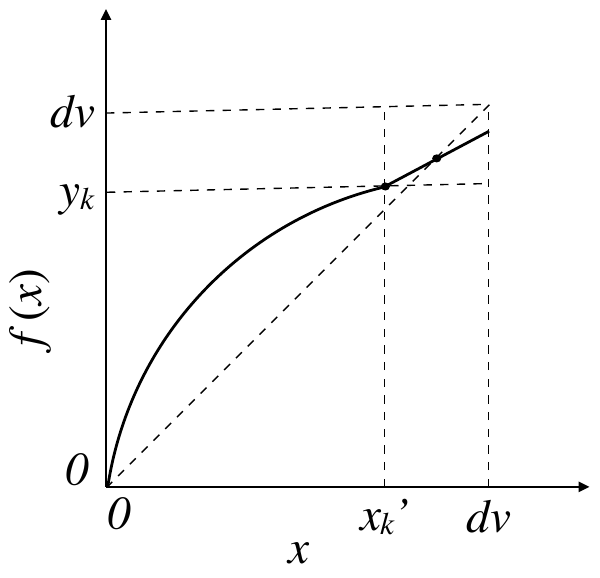}

\protect\caption{Representative plot of the return map on the bottom edge of conduction
state 1 in the asymmetric D-D case. The fixed point corresponding
to anti-phase locking which was at $dv$ in the symmetric case is
shifted inside away from $dv$ in the asymmetric case\label{fig:dd-asym-return-map}}
\end{figure}

\begin{figure}
\includegraphics[scale=0.8]{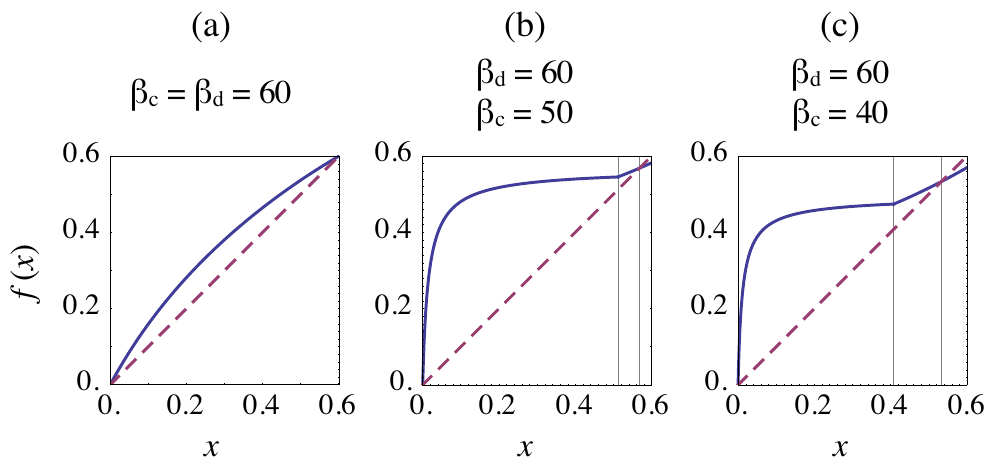}

\protect\caption{Comparison of return maps in the symmetric (a) and asymmetric (b and
c) D-D configurations for constant $\alpha=10$. Both symmetric and
asymmetric configurations have a fixed point at $0$ corresponding
to in-phase locking (which is unstable here as $\beta>\alpha$ condition
is satisfied) along with another fixed point, which in symmetric case,
is at $dv$ (perfect anti-phase locking) but in asymmetric case shifts
away from $dv$ \label{fig:dd-sym-asym-return-maps}}
\end{figure}

\begin{figure}
\includegraphics[scale=0.8]{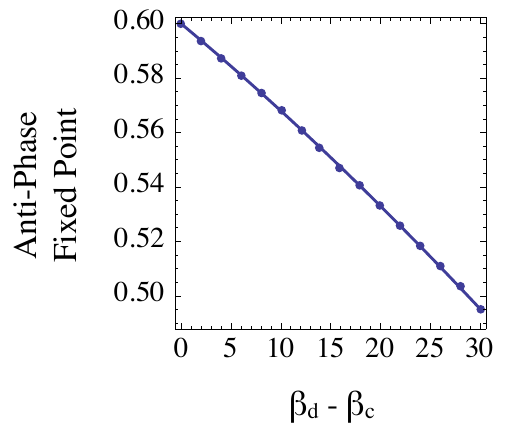}

\protect\caption{Numerical simulations illustrating the fixed point close to $dv$
shifts away from $dv$ with increasing difference between $\beta_{c}$
and $\beta_{d}$ in the asymmetric case\label{fig:dd-asym-shift-fixed-point}}
\end{figure}

\Figref{dd-asym-return-map} shows a representative return map for
the asymmetric D-D configuration. The poincare section chosen in the
symmetric D-D case was the left edge of conduction state 1. Due to
symmetry, the left edge of conduction state 1 is same as the bottom
edge of conduction state 1. Hence the return maps in the symmetric
D-D case can be compared with the return maps in the asymmetric D-D
case as if they were drawn on the same edge. \Figref{dd-sym-asym-return-maps}
shows a comparison of the return maps of a symmetric case $(\beta_{c}=\beta_{d}=60,\,\alpha=10)$
with that of two asymmetric cases $(\beta_{c}=50\, and\,40,\,\beta_{d}=60,\,\alpha=10)$.
The corresponding time domain waveforms and phase plots are shown
in \figref{dd-sym-asym-return-maps}. The figure clearly shows that
the steady state periodic orbit changes from a diagonal (perfect anti-phase
locking) to a butterfly shaped curve (imperfect anti-phase locking)
as the asymmetry increases. However, the time domain waveforms for
butterfly shaped periodic orbits would still be very similar in appearance
to anti-phase locking. The fixed point close to $dv$ in the return
map shifts away from $dv$ as the difference between $\beta_{c}$
and $\beta_{d}$ increases. This trend can be seen in \figref{dd-asym-shift-fixed-point}
which shows the movement of the anti-phase fixed point with $\beta_{d}-\beta_{c}$
for fixed $\beta_{d}=60$ and $\alpha=10$. For $\beta_{c}>\beta_{d}$,
the cycles will be of the type $4\rightarrow2b\rightarrow4$ and $1\rightarrow2c\rightarrow4\rightarrow2a\rightarrow1$,
and the return map will have to be drawn on an edge of state 4. The
return map in this case will be analogous to the $\beta_{c}<\beta_{d}$
case with $\beta_{c}$ and $\beta_{d}$ interchanged.

\begin{figure}
\includegraphics[scale=0.8]{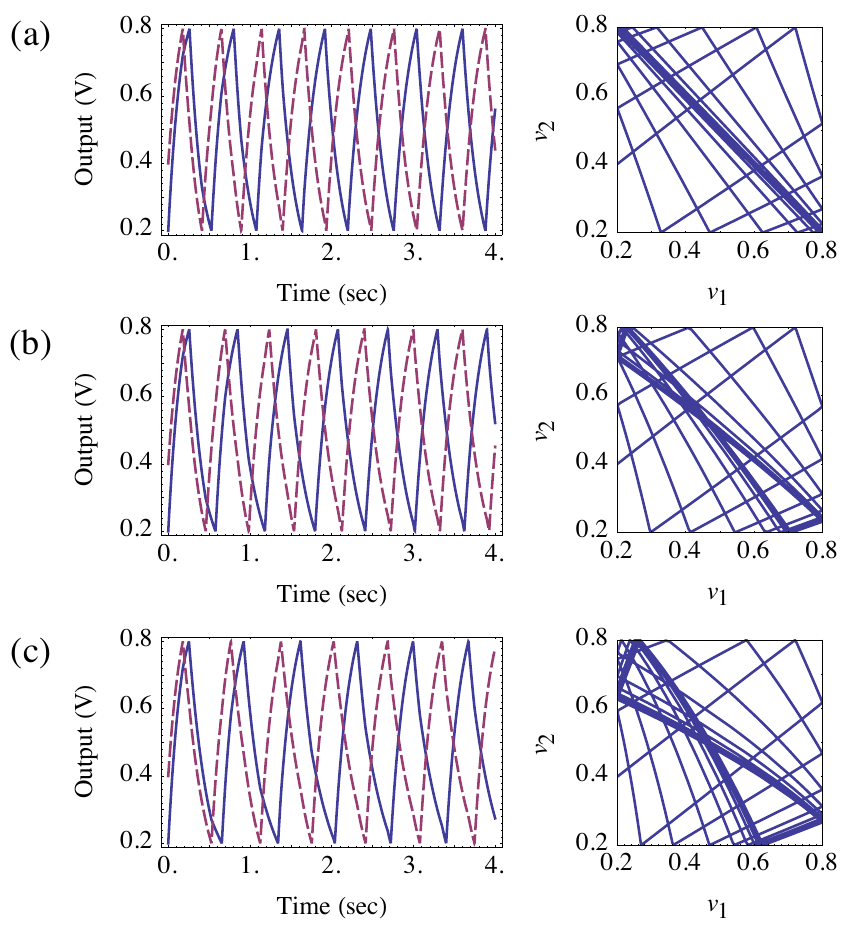}

\protect\caption{Time domain waveforms and phase plots corresponnding to the configurations
in \figref{dd-sym-asym-return-maps}(a, b and c). The steady state
periodic orbits can be seen clearly in the phase plots to transform
from a diagonal (perfect anti-phase locking) in the symmetric case
(a) to a butterfly shaped curve (imperfect anti-phase locking) as
the asymmetry increases and the anti-phase fixed point in the return
map shifts away from $dv$\label{fig:dd-sym-asym-time-phase}}

\end{figure}

\section{D-R coupled oscillator dynamics\label{sec:D-R-coupled-oscillator}}

\begin{figure}
\includegraphics[scale=0.65]{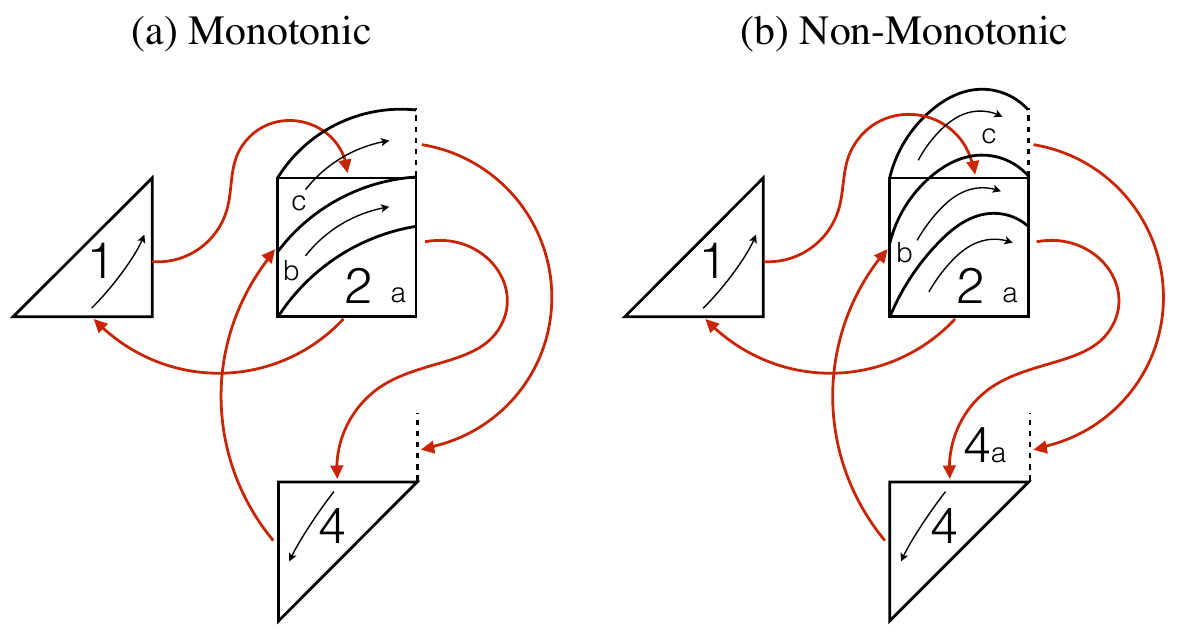}

\protect\caption{Symmetry reduced space in the D-R coupled oscillator system. There
is only a single symmetry due to identical oscillators\label{fig:dr-symmetry-reduced}}
\end{figure}

\begin{figure}
\includegraphics[scale=0.6]{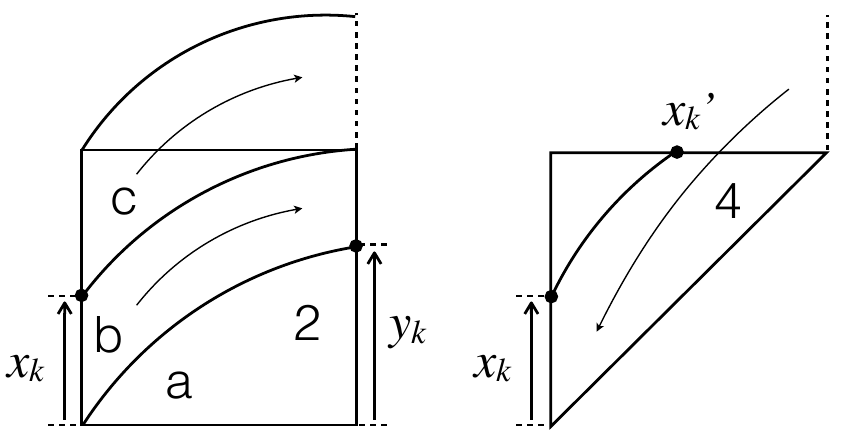}

\protect\caption{Symmetry reduced space for the D-R coupled oscillator system in states
1 and 4 with the definition of $x_{k}$, $x_{k}'$ and $y_{k}$. $f_{4}$
is the mapping from top edge of state 4 to left edge of state 4 and
$f_{2a}$, $f_{2b}$ and $f_{2c}$ are mappings between edges of state
2 as shown\label{fig:dr-xk}}
\end{figure}

In this section we consider the dynamics of a D-R coupled system.
This is of interest because of its ease of fabrication, relaxed conditions
for oscillations and already published reports of such coupled oscillatory
systems \cite{Shukla:2014aa}. We consider coupling of identical oscillators
and hence we define $\beta_{1}=\beta_{2}=\beta$ and $\beta_{s1}=\beta_{s2}=\beta_{s}$.
Unlike the D-D coupled oscillator case, the notion of symmetric charging
and discharging does not apply in D-R coupled oscillator case because
the circuit by construction is different for charging and discharging.
During charging a part of the net charging current charges up the
output capacitor whereas the rest of it flows through the pull-down
resistance to ground. The process of discharging has no such leakage
component. In terms of the conductance ratio $\beta$, this can be
explained by the fact that the net charging component in the matrix
A is ($\beta+\beta_{s}$) and it is always greater than the discharging
component $\beta_{s}$. However, the flows can still be simplified
for analysis as was described in section \ref{sec:Phase-Space,-Flows}.
The simplification assumes that the flows are monotonic in the regions
of operation in all four conduction states, but the direction of monotonicity
is different from the D-D coupled oscillator case as shown in \figref{Transitions}.
For our analysis, a particular type of non-monotonicity is allowed
in state 2 (and state 3) as shown in \figref{dr-symmetry-reduced}.
Here the fixed point for conduction state 2 satisfies the condition
of oscillation shown in \figref{coupled-cond}, but the flow in state
2 as shown in the symmetry reduced space (\figref{dr-symmetry-reduced})
is non monotonic. We will consider the case of identical oscillators
, and following the methodology of the asymmetric D-D case, we can
reduce the symmetry of identical oscillators as shown in \figref{dr-symmetry-reduced}.
In this case, two kinds of cycles are possible - $4\rightarrow2b\rightarrow4$
and $4\rightarrow2c\rightarrow4a\rightarrow4$. To find the fixed
points of the system, we draw the return map on the top edge of conduction
state 4 as the Poincare section. Because it is a symmetry reduced
space, we will have to consider the second return map for cycles of
the type $4\rightarrow2b\rightarrow4$ but only the first return map
for $4\rightarrow2c\rightarrow4a\rightarrow4$ type cycles. Let $f_{4}$
be the mapping from top edge of state 4 to its left edge, $f_{4a}$
be the mapping from the extended right edge of state 4 to its top
edge, $f_{2a}$, $f_{2b}$ and $f_{2c}$ be the mappings between edges
of state 2, and $f_{4}(x_{k}')=x_{k}$ as shown in \figref{dr-xk}.
Then the return map $f$ is given by:
\begin{equation}
f(x)=\begin{cases}
f_{4}\circ f_{2c}\circ f_{4a}(x), & 0\leq x<x_{k}'\\
f_{4}\circ f_{2b}\circ f_{4}\circ f_{2b}(x), & x_{k}'\leq x<dv
\end{cases}\label{eq:dr-return-map}
\end{equation}

\emph{\\Proposition 3} : \emph{Conditions for existence of stable
periodic orbit in D-R coupled oscillator system:} If $f$ is the return
map for the D-R coupled oscillator system on the top edge of state
4 in the symmetry reduced state space (\figref{dr-symmetry-reduced})
then
\begin{enumerate}[label=(\alph*),nosep]
\item $f$ is piece-wise continuous with discontinuity at $x_{k}'$. Moreover,
$f(x_{k+}')=y_{k}$ and $f(x_{k-}')=dv$.
\item $f$ has at least one fixed point in the interval $x_{k}'<x<dv$ at,
say, $x_{f}$
\item If $y_{k}>x_{k}'$, $f$ has at least one stable fixed point in the
interval $x_{k}'<x<dv$\\
\end{enumerate}
\emph{Proof }: (a) The argument is the same as in \emph{Proposition
2.} The return map is separately continuous in intervals $[0,x_{k}')$
and $(x_{k}',dv]$ as it is the composition of continuous flows. From
(\ref{eq:dr-return-map}) we can see that $f(x_{k+}')=y_{k}$ and
$f(x_{k-}')=dv$.

(b) In the interval $x_{k}'<x<dv$, the fixed points of the first
return map $f_{4}\circ f_{2b}$ will also be the fixed points for
its second return map (which is $f$). Now $f_{4}\circ f_{2b}(x_{k}')=dv$
and $f_{4}\circ f_{2b}(dv)=y_{k}$. As $f_{4}\circ f_{2b}$ is continuous
(and hence decreasing) in this interval, there exists a fixed point
for $f_{4}\circ f_{2b}$, and hence $f$, in the interval $x_{k}'<x<dv$.

(c) As $f(x_{k+}')=y_{k}$, $f$ is continuous in the interval $x_{k}'<x<dv$
and $f$ has a fixed point at $x_{f}$ where $x_{k}'<x_{f}<dv$, hence
either the fixed point at $x_{f}$ is stable or there exists another
fixed point in the interval $x_{k}'<x<x_{k}$ which lies in $x_{k}'<x<dv$.
Hence proved.\\

\begin{figure}
\includegraphics[scale=0.8]{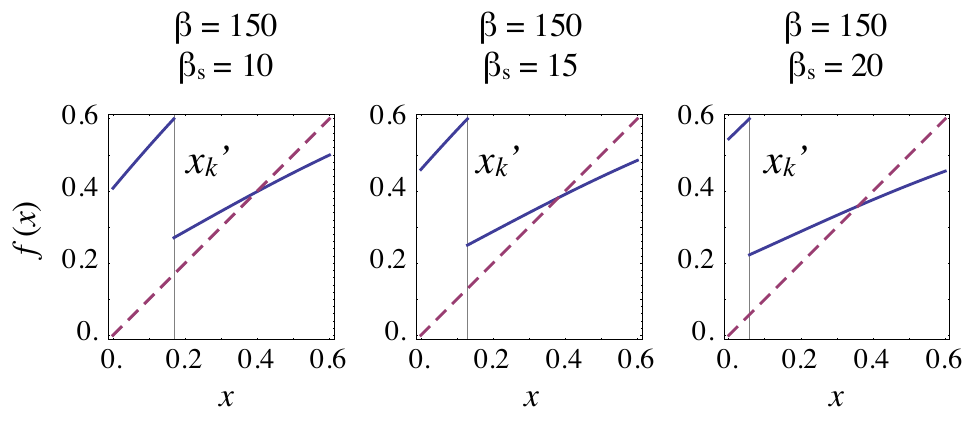}

\protect\caption{Return map on the top edge of state 4 for the D-R coupled oscillator
system for $\alpha=1$, $\beta=150$ and $\beta s$ values of 5, 10
and 20\label{fig:dr-return-maps}}
\end{figure}

\begin{figure}
\includegraphics[scale=0.8]{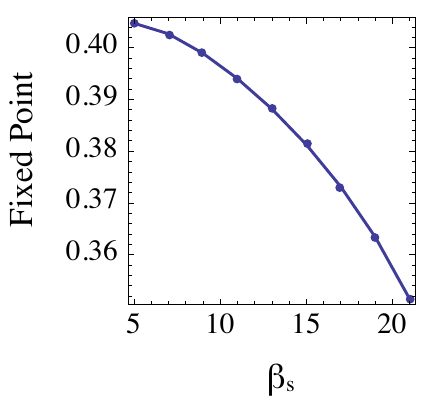}

\protect\caption{Movement of the fixed point $x_{f}$ for fixed $\alpha=1,\,\beta=150$
and varying $\beta_{s}$ for the return map on the top edge of state
4 for the D-R coupled oscillator system\label{fig:dd-fp-var}}
\end{figure}

\Figref{dr-return-maps} shows the return map $f$ on the top edge
of state 4 for the D-R coupled oscillator system for varying $\beta_{s}$.
The return maps in the figure have a single stable fixed point at
$x_{f}$ in the interval $x_{k}'<x<dv$. The movement of the fixed
point $x_{f}$ with $\beta_{s}$ is shown in \figref{dd-fp-var}.

\begin{figure}
\includegraphics[scale=0.8]{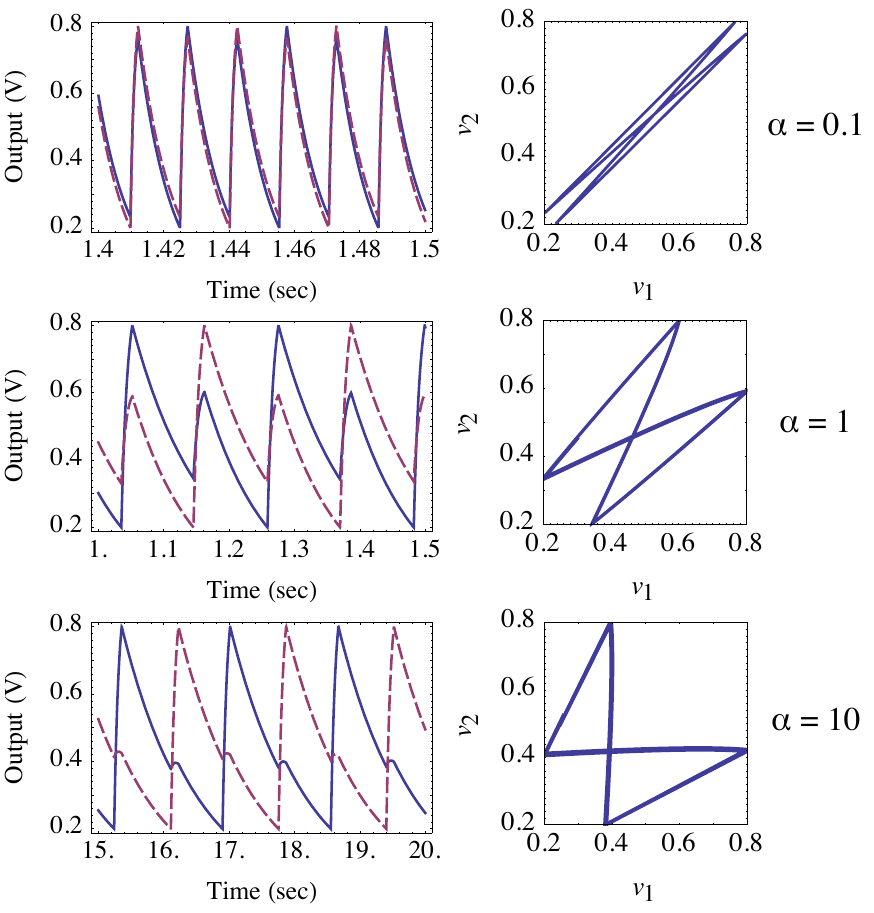}

\protect\caption{Steady state waveforms and phase trajectories for the D-R coupled
oscillator system with $\alpha=0.1$ (top), $\alpha=1$ (middle) and
$\alpha=10$ (bottom). The solid and dashed lines represent the two
oscillators. \label{fig:dr-phase-with-alpha}}
\end{figure}

Another important design consideration for the coupled oscillator
system, is the role of the coupling circuit on the overall system
dynamics, as is seen in \figref{dr-phase-with-alpha}. We note that
as the value of $\alpha$ increases the phase diagram in the $v_{1}\times v_{2}$
plane shows strong sensitivity. In particular, for low values of $\alpha$,
the system shows in-phase locking. As $\alpha$ increases (for intermediate
value of $\alpha$), the butterfly shaped phase plot widens and the
system exhibits a non-monotonic decrease in the output voltages, $v_{1}$
and $v_{2}$ from $v_{h}$ to $v_{l}$. This can also be seen in the
time domain waveforms where the output voltages first decrease to
an intermediate voltage, then increase and again decrease; clearly
demonstrating four possible conduction states (MM, MI, IM and II)
in both phase and time domain plots. Finally, for high values of $\alpha$
the butterfly in the phase plot opens even further, thus making the
decrease of output voltages from $v_{h}$ to $v_{l}$ more monotonic
and the system tends to anti-phase locking, as exhibited in both phase
and time (\figref{dr-phase-with-alpha}).

\section{Experimental verification\label{sec:Experimental-verification}}

An MIT device can be realized using $VO_{2}$ (Vanadium dioxide) which
exhibits unique electronic properties like metal-insulator phase transitions.
$VO_{2}$ has been shown to undergo abrupt first order metal-to-insulator
and insulator-to-metal transitions with upto five orders of change
in conductivity\cite{Ladd:1969aa} and ultra-fast switching times
\cite{Kar:2013aa}. Transitions have been shown to be electrically
driven, thermally driven or a combination thereof. Recent work shows
that for such a transition, a metallic filament structure is formed
which acts as a conduction pathway in the low resistance state of
$VO_{2}$ \cite{Freeman:2013aa}. Also, a series circuit of $VO_{2}$
with a resistive pull down network has been shown to exhibit self-sustained
electrical oscillations\cite{Shukla:2014aa} when conditions of oscillations
as described above are met. Moreover, two such relaxation oscillators
can be electrically coupled to produce synchronized oscillations \cite{Shukla:2014aa}.

\begin{figure}
\includegraphics[scale=0.7]{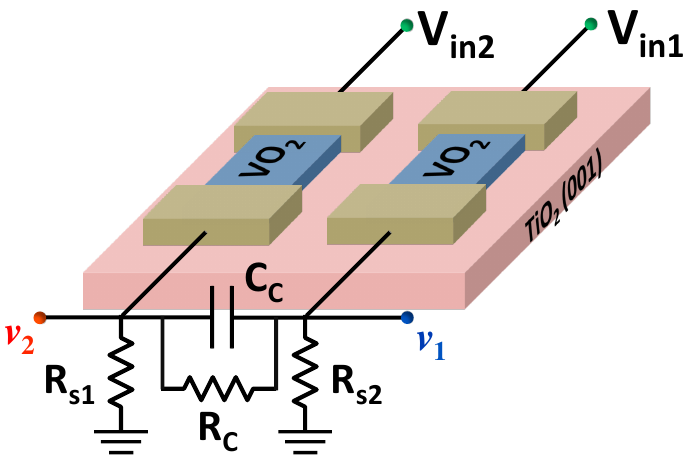}

\protect\caption{Schematic of the experimental setup of coupled $VO_{2}$ oscillators,
with series resistances $R_{s1}$ and $R_{s2}$ respectively, coupled
using a parallel $R_{C}-C_{C}$ circuit\label{fig:exp-circuit}}
\end{figure}

\begin{figure}
\includegraphics[scale=0.7]{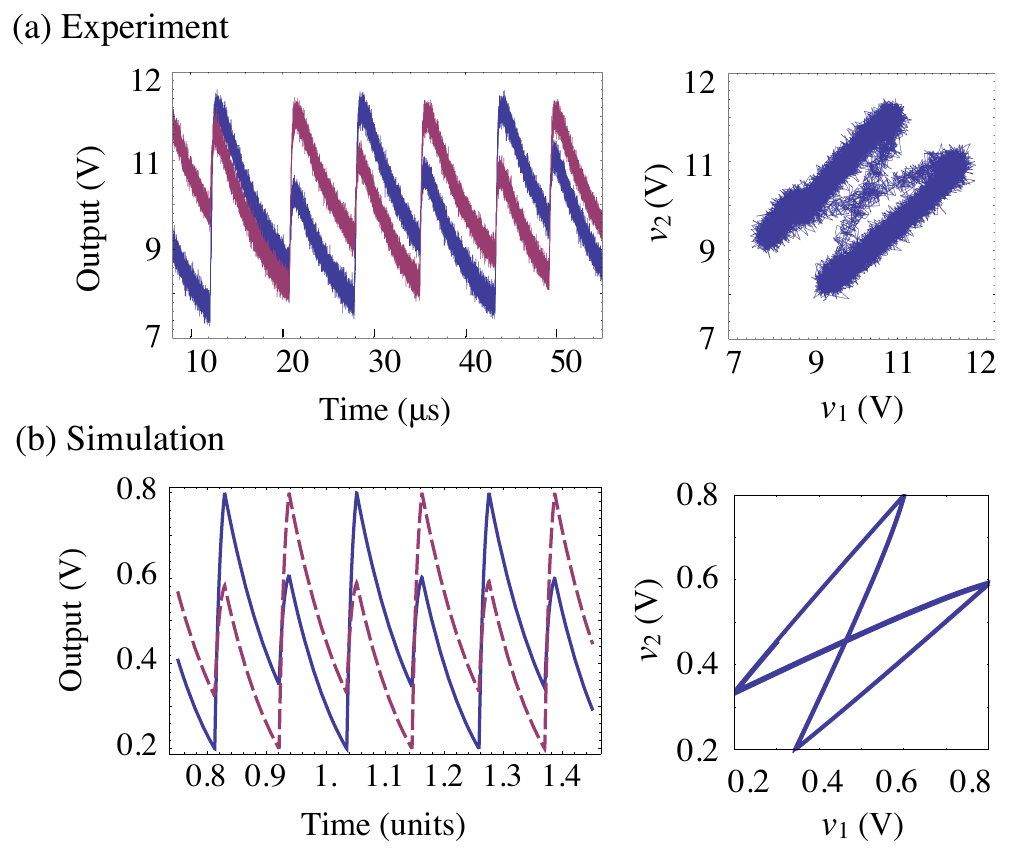}

\protect\caption{Experimental and simulated time domain waveforms in the steady state
and phase plots for a parallel $R_{C}-C_{C}$ coupled oscillator system.
The D-R coupled relaxation oscillator model is used for model development
and simulation. The two waveforms show close match and validate the
model prediction.\label{fig:exp-time-phase}}
\end{figure}

For experimental validation, we apply our models of coupled relaxation
oscillators on a system of two coupled $VO_{2}$ oscillators. \Figref{exp-circuit}
shows a schematic representation of the coupled circuit with a parallel
resistance ($R_{C}$) and capacitance ($C_{C}$) as the coupling circuit.
Frequency domain results of this system have been previously reported
\cite{Shukla:2014aa} showing a close match between experiments and
theoretical results of a D-D model; and are not reproduced here. Using
the D-R model developed in this paper, we obtain close match in the
time-domain and phase plots of the oscillator system as well. With
proper calibration of the system parameters, the D-R model described
above shows very close qualitative match with experimental results.
One such experimental result has been shown in \figref{exp-time-phase}
along with model prediction. This validation of the proposed models
enables further design of experiments. It further models and explains
both qualitative and quantitative the role of the system design parameters
on the rich synchornization dynamics.

\section{Conclusions}

This paper presents a model study of the synchronization dynamics
of a pair of identical and electrically coupled relaxation oscillators
when physically realized using MIT devices. Experimental realization
of such devices\cite{Datta:2014aa,Shukla:2014aa} has motivated the
study of their dynamics, with emphasis on phase synchronization, locking
conditions and potential programmability of the phase relations using
electrical means. We investigate the case of a purely MIT based oscillator
(D-D) and that of a hybrid oscillator composed of an MIT device and
a passive resistance (D-R configuration). We show through numerical
and analytical techniques, validated against experimental results,
the existence of out-of-phase locking (in purely capacitive coupling),
in-phase locking (in purely resistive circuits) and the possibility
of bistable circuits (for intermediate values of R and C). This opens
new paradigms for realizing associative computing networks using coupled
oscillators by enabling model studies of such physically realizable
circuit elements.

\begin{acknowledgments}

NS and SD acknowledge funding from the Office of Naval Research through
award N00014-11-1-0665. SD would also like to acknowledge funding,
in part, from the NSF Expeditions in Computing Award-1317560. AP and
AR would like to acknowledge the generous gift of Intel Corporation
which made this work possible.
\end{acknowledgments}

\bibliographystyle{unsrt}
\bibliography{refs}

\begin{thebibliography}{10}

\bibitem{Datta:2014aa}
Suman Datta, Nikhil Shukla, Matthew Cotter, Abhinav Parihar, and Arijit
  Raychowdhury.
\newblock Neuro inspired computing with coupled relaxation oscillators.
\newblock In {\em Proceedings of the The 51st Annual Design Automation
  Conference on Design Automation Conference}, pages 1--6. ACM, 2014.

\bibitem{Shukla:2014aa}
Nikhil Shukla, Abhinav Parihar, Eugene Freeman, Hanjong Paik, Greg Stone,
  Vijaykrishnan Narayanan, Haidan Wen, Zhonghou Cai, Venkatraman Gopalan, Roman
  Engel-Herbert, et~al.
\newblock Synchronized charge oscillations in correlated electron systems.
\newblock {\em Scientific reports}, 4, 2014.

\bibitem{Dorfler:2012aa}
Florian D{\"o}rfler and Francesco Bullo.
\newblock Exploring synchronization in complex oscillator networks.
\newblock {\em arXiv preprint arXiv:1209.1335}, 2012.

\bibitem{Winfree:1967aa}
Arthur~T Winfree.
\newblock Biological rhythms and the behavior of populations of coupled
  oscillators.
\newblock {\em Journal of theoretical biology}, 16(1):15--42, 1967.

\bibitem{Kuramoto:1975aa}
Yoshiki Kuramoto.
\newblock Self-entrainment of a population of coupled non-linear oscillators.
\newblock In {\em International symposium on mathematical problems in
  theoretical physics}, pages 420--422. Springer, 1975.

\bibitem{Kuramoto:2003aa}
Yoshiki Kuramoto.
\newblock {\em Chemical oscillations, waves, and turbulence}.
\newblock Courier Dover Publications, 2003.

\bibitem{Nikonov:2013aa}
Dmitri~E Nikonov, Gyorgy Csaba, Wolfgang Porod, Tadashi Shibata, Danny Voils,
  Dan Hammerstrom, Ian~A Young, and George~I Bourianoff.
\newblock Coupled-oscillator associative memory array operation.
\newblock {\em arXiv preprint arXiv:1304.6125}, 2013.

\bibitem{Izhikevich:2000aa}
Eugene~M Izhikevich.
\newblock Computing with oscillators.
\newblock 2000.

\bibitem{Mallada:2013aa}
Enrique Mallada and Ao~Tang.
\newblock Synchronization of weakly coupled oscillators: coupling, delay and
  topology.
\newblock {\em Journal of Physics A: Mathematical and Theoretical},
  46(50):505101, 2013.

\bibitem{Acebron:2005aa}
Juan~A Acebr{\'o}n, Luis~L Bonilla, Conrad J~P{\'e}rez Vicente, F{\'e}lix
  Ritort, and Renato Spigler.
\newblock The kuramoto model: A simple paradigm for synchronization phenomena.
\newblock {\em Reviews of modern physics}, 77(1):137, 2005.

\bibitem{Rand:1980aa}
RH~Rand and PJ~Holmes.
\newblock Bifurcation of periodic motions in two weakly coupled van der pol
  oscillators.
\newblock {\em International Journal of Non-Linear Mechanics}, 15(4):387--399,
  1980.

\bibitem{Storti:1982aa}
DW~Storti and RH~Rand.
\newblock Dynamics of two strongly coupled van der pol oscillators.
\newblock {\em International Journal of Non-Linear Mechanics}, 17(3):143--152,
  1982.

\bibitem{Kouda:1982aa}
A~Kouda and S~Mori.
\newblock Mode analysis of a system of mutually coupled van der pol oscillators
  with coupling delay.
\newblock {\em International Journal of Non-Linear Mechanics}, 17(4):267--276,
  1982.

\bibitem{Chakraborty:1988aa}
Tapesh Chakraborty and Richard~H Rand.
\newblock The transition from phase locking to drift in a system of two weakly
  coupled van der pol oscillators.
\newblock {\em International Journal of Non-Linear Mechanics}, 23(5):369--376,
  1988.

\bibitem{Saito:1988aa}
T.~Saito.
\newblock On a coupled relaxation oscillator.
\newblock {\em Circuits and Systems, IEEE Transactions on}, 35(9):1147--1155,
  {Sep} 1988.

\bibitem{Kar:2013aa}
Ayan Kar, Nikhil Shukla, Eugene Freeman, Hanjong Paik, Huichu Liu, Roman
  Engel-Herbert, S.~S.~N. Bharadwaja, Darrell~G. Schlom, and Suman Datta.
\newblock Intrinsic electronic switching time in ultrathin epitaxial vanadium
  dioxide thin film.
\newblock {\em Applied Physics Letters}, 102(7):--, 2013.

\bibitem{Hu:1986aa}
Chia-Lun Hu.
\newblock Self-sustained oscillation in an ${R}_{H}$ - {C} or ${R}_{H}$ - {L}
  circuit containing a hysteresis resistor $r_{H}$.
\newblock {\em Circuits and Systems, IEEE Transactions on}, 33(6):636--641,
  {Jun} 1986.

\bibitem{Ladd:1969aa}
Larry~A. Ladd and William Paul.
\newblock Optical and transport properties of high quality crystals of \{V2O4\}
  near the metallic transition temperature.
\newblock {\em Solid State Communications}, 7(4):425 -- 428, 1969.

\bibitem{Freeman:2013aa}
Eugene Freeman, Greg Stone, Nikhil Shukla, Hanjong Paik, Jarrett~A Moyer,
  Zhonghou Cai, Haidan Wen, Roman Engel-Herbert, Darrell~G Schlom, Venkatraman
  Gopalan, et~al.
\newblock Nanoscale structural evolution of electrically driven insulator to
  metal transition in vanadium dioxide.
\newblock {\em Applied Physics Letters}, 103(26):263109, 2013.

\end{thebibliography}

\end{document}